\documentclass[12pt]{article}
\usepackage[letterpaper, margin=1in]{geometry}

\usepackage{amsmath}
\usepackage{amssymb}
\usepackage{graphicx}
\usepackage{booktabs}
\usepackage{xcolor}

\usepackage{hyperref}

\usepackage{natbib}

\usepackage{subcaption}

\usepackage{algorithm}
\usepackage{algpseudocode}

\DeclareMathOperator*{\argmin}{arg\,min}
\DeclareMathOperator*{\argmax}{arg\,max}
\DeclareMathOperator*{\KL}{KL}
\DeclareMathOperator*{\E}{E}

\DeclareMathOperator*{\upc}{\overset{+}{\approx}}
\newcommand{\R}{\textsf{R}\ }
\newcommand{\pck}[1]{\textsf{#1}}
\newcommand{\trp}{^{\prime}}

\title{Fast Bayesian Basis Selection for Functional Data Representation with Correlated Errors}
\author{Ana Carolina da Cruz$^{1*}$ \and \ Camila P.\ E.\ de Souza$^1$ \and \ Pedro H.\ T.\ O.\ Sousa$^2$ \\  \normalsize $^1$Department of Statistical and Actuarial Sciences, University of Western Ontario, Canada\\
\normalsize$^2$Department of Statistics, Federal University of Paraná, Brazil \\
\normalsize$^*$Corresponding author: adacruz@uwo.ca}
\date{}

\begin{document}

\maketitle

\begin{abstract}
Functional data analysis finds widespread application across various fields. While functional data are intrinsically infinite-dimensional, in practice, they are observed only at a finite set of points, typically over a dense grid. As a result, smoothing techniques are often used to approximate the observed data as functions. In this work, we propose a novel Bayesian approach for selecting basis functions for smoothing one or multiple curves simultaneously. Our method differentiates from other Bayesian approaches in two key ways: (i) by accounting for correlated errors and (ii) by developing a variational Expectation-Maximization (VEM) algorithm, which is faster than Markov chain Monte Carlo (MCMC) methods such as Gibbs sampling. Simulation studies demonstrate that our method effectively identifies the true underlying structure of the data across various scenarios, and it is applicable to different types of functional data. Our VEM algorithm not only recovers the basis coefficients and the correct set of basis functions but also estimates the existing within-curve correlation. When applied to the motorcycle, LIDAR (LIght Detection And Ranging) experiment and Canadian weather datasets, our method demonstrates comparable, and in some cases superior, performance in terms of adjusted $R^2$ compared to regression splines, smoothing splines, least absolute shrinkage and selection operator (LASSO) and Bayesian LASSO. 
Our proposed method is implemented in R and codes are available at \url{https://github.com/acarolcruz/VB-Bases-Selection}.
\end{abstract}


\section{Introduction}

In Functional Data Analysis (FDA), a common problem involves assessing observations from a functional $\mathcal{Y}$ over a continuum $\mathcal{T}$, where observations, denoted ${y}_i(t), i = 1, \dotsc, n$, are referred to as functional data. Since $y(\cdot)$ is not observed at every point, but rather across a dense grid, smoothing techniques are often employed to convert the observed data into functions. This process, referred to as functional data representation, is commonly achieved via basis function expansion methods, using different types of basis functions, such as Fourier or B-splines \citep{ramsay1991, ramsay2005, srivastava2016, kokoszka2017, basna2022}. Comprehensive reviews of functional data analysis methodologies are provided by \cite{goia2016} and \cite{aneiros2019}.

One common method for estimating a smooth function, represented by a linear combination of known basis functions, involves finding the set of coefficients for this linear combination that minimizes a least squares criterion, referred to as regression splines \citep[Chapter 4]{ramsay2005}. Another method, smoothing splines \citep[Chapter 5]{ramsay2005}, finds the coefficients by minimizing a penalized least squares criterion, which includes a penalty term that measures the function’s roughness. The smoothing parameter, which controls the trade-off between data fit and smoothness, is typically determined through cross-validation \citep{craven1978, rice1991, ramsay2005}. Although the smoothing splines method effectively manages smoothness, it does not perform basis selection, which is the process of identifying the most relevant basis functions from a set of candidates.

Bayesian techniques have also significantly advanced the field of functional data representation, providing robust and flexible modelling methods. For example, Gaussian processes, which are non-parametric models, are widely used in Bayesian functional data analysis due to their ability to model and smooth complex data \citep{williams2006}. Additionally, sparsity-inducing prior approaches \citep{kuo1998, rao2005, crainiceanu2009, ghosh2015, yang2024} can be employed for basis function selection, offering an effective approach to distinguish between relevant and irrelevant basis functions. For instance, \cite{sousa2024} utilized sparsity-inducing priors for simultaneous basis function selection across multiple curves by introducing a hierarchical model and sampling from the posterior distribution via a Gibbs sampler, while assuming independent errors. Considering a different perspective, \cite{alves2024} focus on a Bayesian approach for selecting the appropriate number of knots in smoothing splines.

Independence across observations within a curve is often assumed when performing functional data representation, as it greatly simplifies model inference derivation. However, this assumption may not hold true in certain scenarios, especially when working with repeated measurements or longitudinal data \citep{goldsmith2011, huo2023}. As a result, ignoring the data correlation structure can lead to inaccurate inference results \citep{demidenko2013} (see Figure \ref{fig:curve_sim12}). Under a Bayesian framework, introducing a correlation structure within a curve poses a challenge in deriving the posterior distribution for the correlation parameters due to its non-conjugacy in most proposed structure cases. The incorporation of a data correlation structure often requires additional steps within the chosen Bayesian inference algorithm to address this issue. For instance, \cite{dias2013} employed the Metropolis-Hastings method within a Gibbs sampler to account for non-conjugacy when estimating a correlation function decay parameter.

The use of Markov chain Monte Carlo (MCMC) methods, such as the Gibbs sampler, for basis function selection in functional data representation, as shown in \cite{sousa2024}, can be computationally costly, especially in large data settings. Therefore, considering alternative Bayesian inference approaches, such as variational inference, may yield more computationally efficient results \citep{goldsmith2011, earls2017, huo2023, xian2024, alves2024}. Variational inference algorithms, such as variational Bayes (VB) or variational EM \citep{blei2017, li2023, huo2023}, aim to find the so-called variational distribution that best approximates the true posterior instead of sampling from the posterior distribution as MCMC techniques. Additionally, when performing basis selection for functional data representation, it is important to recognize that the assumption of independence across observations within a curve may not hold true in many situations. 

Thus, in this work, we propose a novel Bayesian approach for selecting basis functions for smoothing one or multiple curves simultaneously. Our approach accounts for correlated errors and consists of a variational EM algorithm instead of a MCMC sampling such as Gibbs sampler for faster inference. Since our main interest lies in basis function selection, we focus on obtaining a point estimate for the proposed correlation decay parameter while still approximating the full posterior distributions for the other model parameters. In the E-step of our variational EM algorithm, we fix the correlation decay parameter and estimate the variational distributions for the other model parameters using VB. In the subsequent M-step, we use these VB estimates to maximize the evidence lower bound (ELBO) with respect to the correlation decay parameter. The algorithm iterates until ELBO convergence. As a result, all model parameters are obtained simultaneously and automatically, facilitating for a rapid and adaptive selection of basis functions while incorporating a data correlation structure. Our proposed method is implemented in R and codes are available at \url{https://github.com/acarolcruz/VB-Bases-Selection}.

This paper is organized as follows: Section 2 provides an overview of variational inference, along with a detailed description of our proposed model and VB algorithm. Section 3 presents the results of simulation studies designed to evaluate the performance of our method under various scenarios. In Section 4, we apply our methodology to three real datasets, showcasing its practical utility. Finally, Section 5 offers a conclusion of our study and discusses the implications of our proposed method.

\section{Methods}

\subsection{Variational Bayes}
\label{sec:vb_basis}

Unlike MCMC methods that approximate the exact posterior distribution by sampling from the stationary distribution of the model parameters, variational inference takes on an optimization approach for approximating probability densities \citep{jordan1999, wainwright2008}. Variational Bayes (VB) is a variational inference algorithm that approximates the posterior by another density restricted to a family with some desired properties \citep{blei2017}. The approximated posterior is determined by the density in that family that minimizes the Kullback-Leibler (KL) divergence to the exact posterior. Let $\boldsymbol{\theta}$ be the set of model parameters, $q \in \mathcal{Q}$ be a density function over $\Theta$ (parameter space), and $\boldsymbol{y}$ be the observed data. Then, VB seeks to find a variational distribution $q^*$ in $\mathcal{Q}$ that minimizes the KL divergence to the true posterior, that is, 

\begin{equation*}
q^{*}(\boldsymbol{\theta}) = \argmin_{q \in \mathcal{Q}} \KL(q(\boldsymbol{\theta}) \parallel p(\boldsymbol{\theta} \mid \boldsymbol{y})).
\end{equation*}

The KL divergence between the two densities is defined as:
\begin{align*}
 \KL(q(\boldsymbol{\theta}) \parallel p(\boldsymbol{\theta} \mid \boldsymbol{y})) &= \E\left(\log \frac{q(\boldsymbol{\theta})}{p(\boldsymbol{\theta} \mid \boldsymbol{y})}\right) \nonumber
= \E(\log q(\boldsymbol{\theta})) - \E(\log p(\boldsymbol{\theta} \mid \boldsymbol{y})) \nonumber \\
 &= - \underbrace{\big[ \E(\log p(\boldsymbol{\theta}, \boldsymbol{y}) - \E(\log q(\boldsymbol{\theta}))\big]}_{ELBO} + \log p(\boldsymbol{y}),
\end{align*}
where all expectations are taken with respect to $q(\boldsymbol{\theta})$. Therefore, as an alternative, since $\log p(\boldsymbol{y})$ is a constant with respect to $q(\boldsymbol{\theta})$ and $\KL(q \parallel p)$ is non-negative, we can equivalently maximize the so-called evidence lower bound (ELBO) \citep{goldsmith2011, blei2017}.
Consequently, the task of approximating the posterior density with VB becomes an optimization problem, with the complexity determined by the choice of the density family $\mathcal{Q}$. One widely used family is the mean-field variational family, where the parameters $\boldsymbol{\theta}$ are assumed to be mutually independent, leading to the factorization $q(\boldsymbol{\theta}) = \prod_{i}q_i(\theta_i)$. This assumption ease the analytical computation of $q(\boldsymbol{\theta})$. 

To obtain each variational distribution $q_j(\theta_j)$ we consider the coordinate ascent variational inference (CAVI) algorithm \citep{bishop2006}. Under the CAVI, we optimize the ELBO with respect to a single variational density $q_j(\theta_j)$ while holding the others fixed, resulting in the optimal variational distribution $q_j^*(\theta_j)$ defined as follows:
\begin{equation*}
\log q_j^*(\theta_j) =  E_{q(\boldsymbol{\theta}_{-j})} \Big[\log p(\boldsymbol{y}, \boldsymbol{\theta})\Big]  + \text{constant},
\end{equation*}
where $q(\boldsymbol{\theta}_{-j})$ indicates that the expectation is taken with respect to the variational distributions of all other model parameters except $\theta_j$ and ``constant" represents any terms that do not depend on $\theta_j$.

\subsection{Variational EM}

Suppose that $\theta_l \in \Theta$ is a hyperparameter whose posterior distribution is intractable to obtain due to its non-conjugacy. In variational inference, one approach commonly used to address this issue is the variational EM algorithm \citep{liu2019, li2023, huo2023}. Let $\Lambda \subset \Theta$ represents the set of model parameters for which we can approximate the posterior distribution using VB. The variational EM algorithm iteratively maximizes the ELBO with respect to $q(\Lambda)$ and $\theta_l$ using the following two steps:

\begin{itemize}

\item \textbf{E-step}: Assuming $\theta_l$ fixed, we obtain the optimal variational distribution of $q(\Lambda)$, $q^*(\Lambda)$, using the CAVI algorithm as described in Section \ref{sec:vb_basis}; 

\item \textbf{M-step}: We compute the update of $\theta_l$ as the $\argmax_{\theta_l}$ ELBO$(q^*(\Lambda), \theta_l)$, where the obtained $q^*(\Lambda)$ in the E-step is held fixed.	 
		 
\end{itemize}
 
\subsection{Model}
\label{sec:model}

We consider a Bayesian model framework for basis function selection similar to \cite{sousa2024} and develop a variational inference approach for approximate inference while incorporating a correlation structure into the data. Suppose there are $m$ curves, each with $n_i$ observations at points $t_{ij} \in A \subseteq \mathbb{R}$, where $i \in {1, \ldots, m}$ and $j \in {1, \ldots, n_i}$. Denote the observed value of the $i$-th curve at $t_{ij}$ as $y_{ij}$. We can express $y_{ij}$ as follows:
\begin{equation}
\label{eq:model}
y_{ij} \equiv y_i(t_{ij})= g_i(t_{ij}) + \varepsilon_i(t_{ij}),
\end{equation}
where $g_i(t_{ij})$ is represented as a linear combination of $K$ known basis functions. To perform basis function selection, we further consider a Bernoulli random variable multiplying each coefficient of the expansion so that
\begin{equation}
\label{eq:gi}
g_i(t_{ij}) = \sum_{k=1}^{K} Z_{ki}\beta_{ki}B_k(t_{ij}),
\end{equation}
where the $Z_{ki}$s are Bernoulli independent random variables, with $P(Z_{ki} = 1) = \theta_{ki}$. We define the coefficients from the linear combination of basis functions for the $i$-th functional as the vector $\boldsymbol{\xi}_i = (\xi_{1i}, \cdots, \xi_{Ki})' = (Z_{1i}\beta_{1i}, Z_{2i}\beta_{2i}, \ldots, Z_{Ki}\beta_{Ki})'$. Our approach is commonly used in Bayesian variable selection by methods that make the use of sparsity-inducing priors \citep{kuo1998, rao2005,ghosh2015,sousa2024, he2024}. These priors allow for variable selection by shrinking some coefficients to zero while estimating others with non-zero values. In our context, we incorporate a sparsity-inducing prior in the basis coefficients, $\xi_{ki} = Z_{ki}\beta_{ki}$ in \eqref{eq:gi}, by building a prior that allows the coefficients $\xi_{ki}$ to take zero values with probability $1 - \theta_{ki}$ or being drawn from $\mathcal{N}(0, \tau^2\sigma^2)$ with probability $\theta_{ki}$, allowing for an adaptive basis function selection.

In curve-fitting approaches, it is common to assume that the observations within each individual curve are independent. Specifically, for a given curve $i$, $\varepsilon_i(t_{i1}), \dots, \varepsilon_i(t_{in_i})$ are typically modelled as independent and normally distributed with mean zero and constant variance $\sigma^2$, as in \cite{sousa2024}. However, as previously discussed, depending on the application, this assumption may not hold true  \citep{ramsay2005, goldsmith2011, huo2023}. Therefore, to account for correlation among observations within a curve, we assume $\varepsilon_{i}(\cdot)$ in \eqref{eq:model} follows a Gaussian process with mean zero and covariance functional $\sigma^2\psi(\cdot,\cdot)$. Specifically, we adopt the assumption, as presented in \cite{dias2013}, that $\psi(t, s) = \exp\left(-w|t - s|\right)$, which is the correlation function of an Ornstein–Uhlenbeck process \citep{williams2006}, where $w > 0$ represents the correlation decay parameter. This correlation structure accounts for varying correlations depending on the distance between observations within the same curve; that is, the correlation between $Y_i(t)$ and $Y_i(s)$ decays exponentially as $|t-s|$ increases.  

Based on \eqref{eq:model}, \eqref{eq:gi} and additional assumptions described above, we define the proposed Bayesian hierarchical model as follows:
\begin{align*}
&\boldsymbol{y}_{i} \mid \mathbf{Z}, \boldsymbol{\beta}, \sigma^2; w \sim \mathcal{MVN}\left(\boldsymbol{g}_i, \sigma^2\Psi\right); \nonumber \\
&\beta_{ki} \mid \sigma^2, \tau^2 \sim \mathcal{N}(0, \tau^2\sigma^2); \nonumber \\
&Z_{ki} \mid \boldsymbol{\theta} \sim \mathrm{Bernoulli}(\theta_{ki}); \nonumber \\
&\theta_{ki} \sim \mathrm{Beta}(\mu_{ki}, 1-\mu_{ki});\\
&\tau^2 \sim \mathrm{IG}(\lambda_1, \lambda_2); \nonumber \\
&\sigma^2 \sim \mathrm{IG}(\delta_1, \delta_2),\nonumber
\end{align*}
where $\mathrm{IG}$ denotes the inverse-gamma distribution, $\mathbf{Z} = (\mathbf{Z}_{1}, \dots, \mathbf{Z}_{m})\trp$ with $ \mathbf{Z}_{i} = (Z_{1i}, \dots, Z_{Ki})\trp$, $\boldsymbol{\beta} = (\boldsymbol{\beta}_{1}, \dots, \boldsymbol{\beta}_{m})\trp$ with $ \boldsymbol{\beta}_{i} = (\beta_{1i}, \dots, \beta_{Ki})\trp$,  $\boldsymbol{\theta} = (\theta_{11}, \dots, \theta_{K1},\dots, \theta_{1m}, \dots, \theta_{Km})\trp$, $\boldsymbol{g}_i = (g_i(t_{i1}), \cdots, g_i(t_{in_i}))\trp$ as defined in \eqref{eq:gi} and $\Psi$ is an $n_i \times n_i$ matrix, where the element in the $t_{j}$-th row and $t_{l}$-th column is given by $\exp\left(-w|t_{j} - t_{l}|\right)$, for $t_{j}, t_{l} \in 1, \dots, n_i$, where $w$ controls the decay rate. We adopt a Beta–Bernoulli prior structure for the latent variables $Z_{ki}$ and their corresponding probabilities $\theta_{ki}$, following a standard approach in Bayesian variable selection \citep{kohn2001, liang2023, sousa2024}, which mitigates prior sensitivity on the latent variables. Moreover, we assume a priori dependence between the basis coefficients $\beta_{ki}$ and the error variance $\sigma^2$, a standard conjugate choice previously employed and discussed in \cite{park2008}, \cite{hoff2009}, and \cite{sousa2024}. Although inverse-gamma (IG) priors for variance parameters have been criticized for potential instability in non-informative settings \citep{gelman2006}, they remain widely used due to their conjugacy and analytical tractability \citep{goldsmith2016, sousa2024, xian2024}. In this work, we carefully chose the hyperparameters of the IG prior on the noise variance $\sigma^2$ to reduce instability while allowing the data to guide the estimation.

We develop a variational EM algorithm to infer $\mathbf{Z}, \boldsymbol{\theta}, \boldsymbol{\beta}, \sigma^2, \tau^2$, and $w$. Initially, in the E-step, we treat $w$ as a fixed hyperparameter to derive the variational distributions for the other quantities of interest. Using the mean-field approximation, we assume that the variational distribution factorizes over $\mathbf{Z}, \boldsymbol{\theta}, \boldsymbol{\beta}, \sigma^2,$ and $\tau^2$, enabling a tractable approximation. Subsequently, in the M-step, we update the hyperparameter $w$ by directly maximizing the ELBO. Specifically, given the obtained variational distribution of $\mathbf{Z}, \boldsymbol{\theta}, \boldsymbol{\beta}, \sigma^2,$ and $\tau^2$, we update $w$ by setting the first derivative of the ELBO with respect to $w$ to zero. Since an analytical solution is not available, we use an optimization algorithm to find the optimal value of $w$.

\subsection{Proposed variational EM algorithm}
\label{sec:vb_alg}

In what follows, we derive the variational distribution of all model parameters except $w$ using the CAVI algorithm described in Section \ref{sec:vb_basis}. The update of the estimate of $w$ is then computed by directly maximizing the ELBO with respect to $w$ given the obtained variational distributions. Therefore, in this section we present the E-step of our variational EM algorithm. For the M-step, we use a quasi-Newton method to optimize the value of $w$. Our algorithm is summarized in Algorithm \ref{alg:vem}. 

Assuming $w$ fixed and considering the mean-field approximation, the variational distribution of $\mathbf{Z}, \boldsymbol{\theta}, \boldsymbol{\beta}, \sigma^2,$ and $\tau^2$ can be factored as follows: 

\begin{equation}
\label{vb_factorization}
q(\mathbf{Z}, \boldsymbol{\theta}, \boldsymbol{\beta}, \sigma^2, \tau^2) = \prod_{i=1}^{m}\prod_{k = 1}^{K} q(Z_{ik}) \times  \prod_{i=1}^{m}\prod_{k = 1}^{K} q(\theta_{ik}) \times \prod_{i = 1}^{m} q(\boldsymbol{\beta}_{i}) \times q(\sigma^2) \times q(\tau^2).
\end{equation}

Then, using the CAVI algorithm, we derive an update equation for each quantity in \eqref{vb_factorization} by computing the expectation of the so-called complete-data likelihood with respect to all quantities except the one of interest, where the complete-data likelihood is defined as follows:
\begin{align}
\label{eq:complete_like}
p(\mathbf{Y}, \mathbf{Z}, \boldsymbol{\theta}, \boldsymbol{\beta}, \sigma^2, \tau^2; w) &= p(\mathbf{Y} \mid \mathbf{Z}, \boldsymbol{\theta}, \boldsymbol{\beta}, \sigma^2, \tau^2; w) p(\mathbf{Z}, \boldsymbol{\theta}, \boldsymbol{\beta}, \sigma^2, \tau^2) \nonumber \\
&= p(\mathbf{Y} \mid \mathbf{Z}, \boldsymbol{\beta}, \sigma^2; w) p(\mathbf{Z} \mid \boldsymbol{\theta}) p(\boldsymbol{\beta} \mid  \sigma^2, \tau^2)p(\boldsymbol{\theta})p(\sigma^2)p(\tau^2).
\end{align}

Then, for instance, the optimal variational distribution for $Z_{ki}$, $q^*(Z_{ki})$, is given by
\begin{equation*}
\log q^*(Z_{ki}) = E_{q(-Z_{ki})}(\log p(\mathbf{Y}, \mathbf{Z}, \boldsymbol{\theta}, \boldsymbol{\beta}, \sigma^2, \tau^2; w)) + \text{constant},
\end{equation*}
where $q(-Z_{ki})$ indicates that the expectation is taken with respect to the variational distributions of all other random variables except $Z_{ki}$, and ``constant" represents all terms that do not depend on $Z_{ki}$. Next, in Section \ref{sec:vb_eq}, we derive the update equation for each quantity of interest. The expectations shown in the update equations are defined in Section \ref{sec:vb_expectations}. In Section \ref{sec:vem_elbo} we derive the ELBO used as the stop criterion in our VB algorithm.

\subsubsection{VB update equations}
\label{sec:vb_eq}
\textbf{• Update equation for $q(Z_{ki}).$}
\begin{align}
\label{eq_qZ}
\log q^*(Z_{ki}) &= E_{q(-Z_{ki})}(\log p(\mathbf{Y}, \mathbf{Z}, \boldsymbol{\theta}, \boldsymbol{\beta}, \sigma^2, \tau^2; w)) + \text{constant} \nonumber \\
&\upc E_{q(-Z_{ki})}(\log p(\mathbf{Y} \mid \mathbf{Z}, \boldsymbol{\beta}, \sigma^2; w)) + E_{q(-Z_{ki})}(\log p(\mathbf{Z} \mid \boldsymbol{\theta})).
\end{align}

Taking the expectations in \eqref{eq_qZ}, the terms that do not depend on $Z_{ki}$ will be added to the constant term. For convenience, we use $\overset{+}{\approx}$ to denote equality up to a constant. Hence, 
\begin{align}
\log q^*(Z_{ki}) \upc \sum_{r=0}^{1} \mathrm{I}(Z_{ki} = r) & \Big\{-\frac{n_i}{2}E_{q(\sigma^2)}\log(\sigma^2) \nonumber \\
&- \frac{1}{2}E_{q(\sigma^2)}\left(\frac{1}{\sigma^2}\right)E_{q(\mathbf{Z}_{-ki})\cdot q(\boldsymbol{\beta})}\big[(\mathbf{y}_i - \boldsymbol{\tilde g}_i)'\Psi^{-1}(\mathbf{y}_i - \boldsymbol{\tilde g}_i)\big] \nonumber \\
& + r E_{q(\theta_{ki})} \log(\theta_{ki}) + (1 - r) E_{q(\theta_{ki})} \log( 1 - \theta_{ki})\Big\}, \nonumber
\end{align}
where $q(\mathbf{Z}_{-ki})$ represents the variational distribution of the other entries in $\mathbf{Z}_i$, except for $Z_{ki}$. Additionally, $\boldsymbol{\tilde g}_i = \tilde G_{i}'\boldsymbol{\beta}_i$, with
$$ \tilde{G_{i}} = (\tilde{\mathbf{G}}_{i1} \cdots \tilde{\mathbf{G}}_{ij} \cdots \tilde{\mathbf{G}}_{in_i})_{K \times n_i},$$ 
and $\tilde{\mathbf{G}}_{ij} = (Z_{1i}B_1(t_{ij}), \dots, Z_{{k-1}i}B_{k-1}(t_{ij}), rB_k(t_{ij}), Z_{{k+1}i}B_{k+1}(t_{ij}), \dots, Z_{Ki}B_{K}(t_{ij}))'$.

Therefore, $q^*(Z_{ki})$ has a Bernoulli distribution with parameter
\begin{equation}
\label{eq:pki}
p^*_{ki} = \frac{\exp{(\alpha_{ki1}})}{\sum_{r = 0}^{1}\exp{(\alpha_{kir}})},
\end{equation}
where 
\begin{align}
\alpha_{kir} = & -\frac{n_i}{2}E_{q(\sigma^2)}\log(\sigma^2) = - \frac{1}{2}E_{q(\sigma^2)}\left(\frac{1}{\sigma^2}\right)E_{q(\mathbf{Z}_{-ki})\cdot q(\boldsymbol{\beta})}\big[(\mathbf{y}_i - \boldsymbol{\tilde g}_i)'\Psi^{-1}(\mathbf{y}_i - \boldsymbol{\tilde g}_i)\big] \nonumber \\
&+ r E_{q(\theta_{ki})} \log(\theta_{ki}) + (1 - r) E_{q(\theta_{ki})} \log( 1 - \theta_{ki}). \nonumber
\end{align}

\textbf{• Update equation for $q(\theta_{ki}).$} Since only $p(\mathbf{Z} \mid \boldsymbol{\theta})$ and $p(\boldsymbol{\theta})$ in \eqref{eq:complete_like} depend on ${\theta_{ki}}$, we derive $q^*(\theta_{ki})$ as follows:
\begin{align}
\log q^*(\theta_{ki}) &\upc E_{q(-\theta_{ki})}(\log p(\mathbf{Z} \mid \boldsymbol{\theta})) + E_{q(-\theta_{ki})}(\log p(\boldsymbol{\theta})) \nonumber \\
& \upc E_{q(-\theta_{ki})}(\log p({Z}_{ki} \mid \theta_{ki})) + E_{q(-\theta_{ki})}(\log p(\theta_{ki})) \nonumber \\
& = E_{q(Z_{ki})}Z_{ki}\log\theta_{ki} +  E_{q(Z_{ki})}(1 - Z_{ki})\log(1 - \theta_{ki}) \nonumber \\
& + (\mu_{ki} - 1)\log\theta_{ki} + (1 - \mu_{ki} - 1)\log(1-\theta_{ki}) \nonumber \\
&= (E_{q(Z_{ki})}Z_{ki} + \mu_{ki} - 1)\log\theta_{ki} + (2 - E_{q(Z_{ki})}Z_{ki} - \mu_{ki} - 1)\log(1 - \theta_{ki}) \nonumber .
\end{align}

Hence, $q^*(\theta_{ki})$ is a Beta distribution with parameters
\begin{equation}
\label{eq_theta_a1}
a_{ki1} = E_{q(Z_{ki})}Z_{ki} + \mu_{ki}
\end{equation}
and
\begin{equation}
\label{eq_theta_a2}
a_{ki2} = 2 - E_{q(Z_{ki})}Z_{ki} - \mu_{ki}
\end{equation}

\textbf{• Update equation for $q(\boldsymbol{\beta}_i).$} Considering that only the terms $p(\mathbf{Y} \mid \mathbf{Z}, \boldsymbol{\beta}, \sigma^2; w)$ and $p(\boldsymbol{\beta} \mid  \sigma^2, \tau^2)$ in \eqref{eq:complete_like} depend on $\boldsymbol{\beta}_i$, we derive the $q^*(\boldsymbol{\beta}_i)$ as follows:
\begin{align}
\label{eq_qbeta}
\log q^*(\boldsymbol{\beta}_i) & \upc E_{q(-\boldsymbol{\beta}_i)}(\log p(\mathbf{Y} \mid \mathbf{Z},  \boldsymbol{\beta}, \sigma^2; w)) + E_{q(-\boldsymbol{\beta}_i)}(\log p(\boldsymbol{\beta} \mid  \sigma^2, \tau^2)) \nonumber \\
& \upc -\frac{n_i}{2}E_{q(\sigma^2)}\log(\sigma^2) - \frac{1}{2}E_{q(\sigma^2)}\left(\frac{1}{\sigma^2}\right)E_{q(\mathbf{Z})}\left[(\mathbf{y}_i - \boldsymbol{g}_i)'\Psi^{-1}(\mathbf{y}_i - \boldsymbol{ g}_i)\right] \nonumber \\
& - \frac{K}{2}\left(E_{q(\sigma^2)}[\log(\sigma^2)] + E_{q(\tau^2)}[\log(\tau^2)]\right) - \frac{1}{2}E_{q(\sigma^2)}\left(\frac{1}{\sigma^2}\right)E_{q(\tau^2)}\left(\frac{1}{\tau^2}\right)\boldsymbol{\beta}_i'\boldsymbol{\beta}_i. \nonumber
\end{align}

Let $G_{i}$ be a $K \times n_i$ matrix with column vectors $\mathbf{G}_{ij} = (Z_{1i}B_1(t_{ij}), \dots, Z_{Ki}B_K(t_{ij}))'$. Hence, $E_{q^*(\mathbf{Z})} \mathbf{G}_{ij} = (p^*_{1i}B_1(t_{ij}), \dots, p^*_{Ki}B_K(t_{ij}))',$ where $p^*_{ki}$ is defined as in \eqref{eq:pki}. 

Therefore, by completing the squares with respect to $\boldsymbol{\beta}_i$ we have that $q^*(\boldsymbol{\beta}_i)$ is a multivariate normal distribution with parameters
\begin{equation}
\label{eq_qbeta_mu}
\mu_{\boldsymbol{\beta}_i} = \left(E_{q(\sigma^2)}\left(\frac{1}{\sigma^2}\right) \mathbf{y}_i'\Psi^{-1}E_{q(\mathbf{Z})} \left[{G}_{i}'\Sigma_{\boldsymbol{\beta}_i}\right]\right)'
\end{equation}
and
\begin{equation}
\label{eq_qbeta_sig}
\Sigma_{\boldsymbol{\beta}_i} = \left[E_{q(\sigma^2)}\left(\frac{1}{\sigma^2}\right)\left(E_{q(\tau^2)}\left(\frac{1}{\tau^2}\right)\mathrm{I}_{K} + E_{q(\mathbf{Z})} \left[G_{i}\Psi^{-1}\right]E_{q(\mathbf{Z})} \left[G_i'\right] \right)\right]^{-1},
\end{equation}
where $\mathrm{I}_{K}$ is the identity matrix of size $K$.

\textbf{• Update equation for $q(\sigma^2).$}
\begin{align}
\label{eq_qsigma2}
\log q^*(\sigma^2) & \upc E_{q(-\sigma^2)}(\log p(\mathbf{Y} \mid \mathbf{Z},  \boldsymbol{\beta}, \sigma^2; w)) + E_{q(-\sigma^2)}(\log p(\sigma^2)) \nonumber \\
& \upc -\left(\frac{\sum_{i = 1}^{m}n_i}{2} + \frac{mK}{2} + \delta_1 + 1 \right) \log(\sigma^2) 
-\Bigg\{ \sum_{i = 1}^{m}E_{q(\mathbf{Z}_i)\cdot q(\boldsymbol{\beta}_i)}\left(\frac{(\mathbf{y}_i - \boldsymbol{g}_i)'\Psi^{-1}(\mathbf{y}_i - \boldsymbol{g}_i)}{2}\right) + \nonumber \\
& E_{q(\tau^2)}\left(\frac{1}{\tau^2}\right) \sum_{i = 1}^{m} E_{q(\boldsymbol{\beta}_i)}\left[\boldsymbol{\beta}_i'\boldsymbol{\beta}_i\right] + \delta_2\Bigg\}\left(\frac{1}{\sigma^2}\right). \nonumber 
\end{align}

Therefore, $q^*(\sigma^2)$ is an inverse-gamma distribution with parameters
\begin{equation}
\label{eq_sigma2_d1}
\delta^*_1 = \frac{\sum_{i = 1}^{m}n_i  + mK + 2\delta_1}{2}
\end{equation}
and
\begin{equation}
\label{eq_sigma2_d2}
\delta^*_2 = \frac{1}{2}\left(\sum_{i = 1}^{m}E_{q(\mathbf{Z}_i)\cdot q(\boldsymbol{\beta}_i)}\left[(\mathbf{y}_i - \boldsymbol{g}_i)'\Psi^{-1}(\mathbf{y}_i - \boldsymbol{ g}_i)\right] +  E_{q(\tau^2)}\left(\frac{1}{\tau^2}\right) \sum_{i = 1}^{m} E_{q(\boldsymbol{\beta}_i)}[\boldsymbol{\beta}_i'\boldsymbol{\beta}_i] + 2\delta_2\right).
\end{equation}

\textbf{• Update equation for $q(\tau^2).$} Since only the terms $p(\boldsymbol{\beta} \mid  \sigma^2, \tau^2)$ and $p(\tau^2) $ in \eqref{eq:complete_like} depend on $\tau^2$, we derive the $q^*(\tau^2)$ as follows:
\begin{align}
\label{eq_qtau2}
\log q^*(\tau^2) & \upc E_{q(-\tau^2)}(\log p(\boldsymbol{\beta} \mid  \sigma^2, \tau^2)) + E_{q(-\tau^2)}(\log p(\tau^2)) \nonumber \\
& \upc -\left(\frac{mK}{2} + \lambda_1 + 1 \right) \log(\tau^2) -\frac{1}{2}\left(2\lambda_2 + E_{q(\sigma^2)}\left(\frac{1}{\sigma^2}\right)\sum_{i = 1}^{m} E_{q(\boldsymbol{\beta}_i)}[\boldsymbol{\beta}_i'\boldsymbol{\beta}_i]\right) \left(\frac{1}{\tau^2}\right). \nonumber 
\end{align}

Similar to $q^*(\sigma^2)$, $q^*(\tau^2)$ is also an inverse-gamma distribution with parameters
\begin{equation}
\label{eq_tau2_l1}
\lambda^*_1 = \frac{mK + 2\lambda_1}{2}
\end{equation}
and
\begin{equation}
\label{eq_tau2_l2}
\lambda^*_2 =  \frac{1}{2}\left(2\lambda_2 + E_{q(\sigma^2)}\left(\frac{1}{\sigma^2}\right)\sum_{i = 1}^{m} E_{q(\boldsymbol{\beta}_i)}\left[\boldsymbol{\beta}_i'\boldsymbol{\beta}_i\right]  \right).
\end{equation}

\subsubsection{Expectations}
\label{sec:vb_expectations}

In this section, we define the expectations in the update equations for the variational distributions derived in Section \ref{sec:vb_eq}. Based on the properties of each type of probability distribution, we obtain that
\begin{equation*}
E_{q(Z_{ki})}[Z_{ki}] = p^*_{ki},
\end{equation*}

\begin{equation*}
E_{q(\boldsymbol{\beta}_i)}[\beta_{ki}^2] =  \mathrm{Var}_{\beta_{ki}} + \mu_{\beta_{ki}}^2,
\end{equation*}

\begin{equation*}
E_{q(\theta_{ki})}[\log\theta_{ki}] = \gamma(a_{ki1}) - \gamma(a_{ki1} + a_{ki2}),
\end{equation*}

\begin{equation*}
E_{q(\theta_{ki})}[\log(1-\theta_{ki})] = \gamma(a_{ki2}) - \gamma(a_{ki1} + a_{ki2} ),
\end{equation*}

\begin{equation*}
E_{q(\sigma^2)}[\log\sigma^2] = \log\delta^*_2 - \gamma(\delta^*_1), \; \;
E_{q(\tau^2)}[\log\tau^2] = \log\lambda^*_2 - \gamma(\lambda^*_1),
\end{equation*}

\begin{equation*}
E_{q(\sigma^2)}\left(\frac{1}{\sigma^2}\right) = \frac{\delta^*_1}{\delta^*_2}, \; \mbox{and} \;
E_{q(\tau^2)}\left(\frac{1}{\tau^2}\right) = \frac{\lambda^*_1}{\lambda^*_2},
\end{equation*}
\noindent where $\gamma(x) = \frac{d}{dx}\Gamma(x)$, where $\Gamma(x)$ is the gamma function and $\gamma(x)$ is the digamma function.

Additionally, using the result $E(X'AX) = \text{trace}(A\text{Var}X) + E(X)'AE(X)$ \citep{seber2012}, we obtain that
\begin{align*}
&E_{q(\mathbf{Z}_i)\cdot q(\boldsymbol{\beta}_i)}\left[(\mathbf{y}_i - \boldsymbol{g}_i)'\Psi^{-1}(\mathbf{y}_i - \boldsymbol{g}_i)\right] \nonumber \\
& = \left(\mathbf{y}_i - E_{q(\mathbf{Z})} [G_{i}'\mu_{\boldsymbol{\beta}_i}]\right)'\Psi^{-1}\left(\mathbf{y}_i - E_{q(\mathbf{Z})}\left[G_{i}'\mu_{\boldsymbol{\beta}_i}\right]\right) + \text{trace}\left(\Psi^{-1}\text{Var}\left[G_i'\boldsymbol{\beta}_i\right]\right),
\end{align*}
where 
\begin{equation*}
\mathrm{Var}\left[G_i'\boldsymbol{\beta}_i\right] = B(\mathbf{t}_i) \text{Var}(\mathbf{Z}_i \circ \boldsymbol{\beta}_i)B(\mathbf{t}_i)',
\end{equation*}
where $B(\mathbf{t}_i)$ is the matrix with the basis functions applied to the vector of evaluation points $\mathbf{t}_i$ for functional $i$ and $\circ$ defines element-wise multiplication. Therefore,
\begin{equation*}
\mathrm{Var}({\boldsymbol{Z}}_{i} \circ \boldsymbol{\beta}_{i}) = \mathrm{Var}({\boldsymbol{Z}}_i)\Sigma_{\boldsymbol{\beta}_i} + \mathrm{Var}({\boldsymbol{Z}}_i)\boldsymbol{\mu}_{\boldsymbol{\beta}_i}\boldsymbol{\mu}_{\boldsymbol{\beta}_i}' + \Sigma_{\boldsymbol{\beta}_i}\mathrm{E}({\boldsymbol{Z}}_i)\mathrm{E}({\boldsymbol{Z}}_i)'.
\end{equation*}

Similarly, $E_{q(\mathbf{Z}_{-k})\cdot q(\boldsymbol{\beta})}\left[(\mathbf{y}_i - \boldsymbol{\tilde g}_i)'\Psi^{-1}(\mathbf{y}_i - \boldsymbol{\tilde g}_i)\right]$ in  \eqref{eq:pki} is obtained as follows:
\begin{align*}
&E_{q(\mathbf{Z}_{-ki})\cdot q(\boldsymbol{\beta})}\left[(\mathbf{y}_i - \boldsymbol{\tilde g}_i)'\Psi^{-1}(\mathbf{y}_i - \boldsymbol{\tilde g}_i)\right]\nonumber \\
& = \left(\mathbf{y}_i - E_{q(\mathbf{Z}_{-ki})} \left[\tilde{G}_{i}'\mu_{\boldsymbol{\beta}_i}\right]\right)'\Psi^{-1}\left(\mathbf{y}_i - E_{q(\mathbf{Z}_{-ki})} \left[\tilde{G}_{i}'\mu_{\boldsymbol{\beta}_i}\right]\right) + \text{trace}\left(\Psi^{-1}\text{Var}\left[\tilde{G}_i'\boldsymbol{\beta}_i\right]\right),
\end{align*}
where 
\begin{equation*}
\mathrm{Var}(\tilde{G}_i'\boldsymbol{\beta}_i) = B(\mathbf{t}_i) \text{Var}(\mathbf{\tilde{Z}}_i \circ \boldsymbol{\beta}_i)B(\mathbf{t}_i)',
\end{equation*}
$\mathbf{\tilde{Z}}_i$ is the vector $\mathbf{{Z}}_i$ with the $k$-th entry equals to a constant $r$ (0, 1) and 
\begin{equation*}
\text{Var}(\mathbf{\tilde{Z}}_i \circ \boldsymbol{\beta}_i) = \mathrm{Var}({\boldsymbol{\tilde Z}}_i)\Sigma_{\boldsymbol{\beta}_i} + \mathrm{Var}({\boldsymbol{\tilde Z}}_i)\boldsymbol{\mu}_{\boldsymbol{\beta}_i}\boldsymbol{\mu}_{\boldsymbol{\beta}_i}' + \Sigma_{\boldsymbol{\beta}_i}\mathrm{E}({\boldsymbol{\tilde Z}}_i)\mathrm{E}({\boldsymbol{\tilde Z}}_i)'. 
\end{equation*}

\subsubsection{Evidence Lower Bound (ELBO)}
\label{sec:vem_elbo}

In this section we derive the ELBO, which is used in the VB algorithm as the convergence criterion. The ELBO is defined as 
\begin{equation*}
\text{ELBO}(q) = E_{q}(\log p(\mathbf{Y}, \mathbf{Z}, \boldsymbol{\theta}, \boldsymbol{\beta}, \sigma^2, \tau^2; w)) - E_{q}(\log q( \mathbf{Z}, \boldsymbol{\theta}, \boldsymbol{\beta}, \sigma^2, \tau^2)).
\end{equation*}

Using the decomposition of the complete-data likelihood given in \eqref{eq:complete_like} and the fact that the variational distribution is assumed to be part of the mean-field family, we calculate the ELBO as follows:
\begin{align*}
\text{ELBO}(q) = & E_{q}(\log p(\mathbf{Y} \mid \mathbf{Z}, \boldsymbol{\beta}, \sigma^2; w)) + 
 E_{q}(\log p(\mathbf{Z} \mid \boldsymbol{\theta})) -  E_{q}(\log q( \mathbf{Z})) +\nonumber \\
& E_{q}(\log p(\boldsymbol{\beta} \mid \sigma^2, \tau^2))
-  E_{q}(\log q(\boldsymbol{\beta}) ) +  E_{q}(\log p(\boldsymbol{\theta})) -  E_{q}(\log q(\boldsymbol{\theta})) +  \\
& E_{q}(\log p(\sigma^2)) -  E_{q}(\log q(\sigma^2) )+  E_{q}(\log p(\tau^2)) -  E_{q}(\log q(\tau^2) ) \nonumber. 
\end{align*}

Consequently, we define each term as follows:
\begin{align}
\label{eq:elbo_1}
&E_{q}(\log p(\mathbf{Z} \mid \boldsymbol{\theta})) -  E_{q}(\log q( \mathbf{Z})) = \\
&= \sum_{i=1}^{m}\sum_{k = 1}^{K}\Big\{p^*_{ki}E_{q(\theta_{ki})}[\log(\theta_{ki})]  + (1 - p^*_{ki})E_{q(\theta_{ki})}[\log(1-\theta_{ki})]  - p^*_{ki}\log (p^*_{ki}) - (1-p^*_{ki})\log(1-p^*_{ki}) \Big\},\nonumber 
\end{align}

\begin{align}
\label{eq:elbo_2}
&E_{q}(\log p(\boldsymbol{\beta} \mid \sigma^2, \tau^2)) -  E_{q}(\log q(\boldsymbol{\beta}) = \\
&= \sum_{i=1}^{m} \Big\{-\frac{K}{2}\left[E_{q(\sigma^2)}\log (\sigma^2) + E_{q(\tau^2)}\log (\tau^2)\right]  + \frac{\log |\Sigma_{\boldsymbol{\beta}_i}|}{2} + \frac{K}{2} \nonumber \\ 
& - \frac{1}{2}E_{q(\sigma^2)}\left(\frac{1}{\sigma^2}\right)E_{q(\tau^2)}\left(\frac{1}{\tau^2} \right)(\text{trace}(\Sigma_{\boldsymbol{\beta}_i}) + \boldsymbol{\mu}_{\boldsymbol{\beta}_i}'\boldsymbol{\mu}_{\boldsymbol{\beta}_i}) \Big\},\nonumber 
\end{align}

\begin{align}
\label{eq:elbo_3}
&E_{q^*}(\log p(\boldsymbol{\theta})) -  E_{q^*}(\log q(\boldsymbol{\theta})) = \\
& \sum_{i=1}^{m}\sum_{k = 1}^{K} \Big\{ (\mu_{ki} - a_{ki1})E_{q^*(\theta_{ki})}[\log(\theta_{ki})] + (1 - \mu_{ki} + a_{ki2})E_{q^*(\theta_{ki})}[\log(1 - \theta_{ki})] \nonumber \\ 
& - \log\Gamma(\mu_{ki})\Gamma(1-\mu_{ki}) + \frac{\log\Gamma(a_{ki1})\Gamma(a_{ki2})}{2} \Big\},  \nonumber 
\end{align}

\begin{align}
\label{eq:elbo_4}
& E_{q^*}(\log p(\sigma^2)) -  E_{q^*}(\log q(\sigma^2) ) =  \\
& =  \delta_1\log\delta_2 - \log\Gamma(\delta_1) - \delta^*_1\log\delta^*_2 + \log\Gamma(\delta^*_1)  +\nonumber  \\
&(\delta^*_1 - \delta_1)E_{q^*(\sigma^2)}\log \sigma^2 + (\delta^*_2 - \delta_2)E_{q^*(\sigma^2)}\left(\frac{1}{\sigma^2}\right), \nonumber
\end{align}

\begin{align}
\label{eq:elbo_5}
&E_{q^*}(\log p(\tau^2)) -  E_{q^*}(\log q(\tau^2) ) =\\
& \lambda_1\log\lambda_2 - \log\Gamma(\lambda_1) - \lambda^*_1\log\lambda^*_2 +\log\Gamma(\lambda^*_1)  +  \nonumber \\ 
& (\lambda^*_1 - \lambda_1)E_{q^*(\tau^2)}\log \tau^2 + (\lambda^*_2 - \lambda_2)E_{q^*(\tau^2)}\left(\frac{1}{\lambda^2}\right), \nonumber 
\end{align}
and
\begin{align}
\label{eq:elbo_6}
&E_{q^*}(\log p(\mathbf{Y} \mid \mathbf{Z}, \boldsymbol{\beta}, \sigma^2; w) =  \\
&= \sum_{i=1}^{m} \Big\{-\frac{n_i}{2}\left(\log 2\pi + E_{q^*(\sigma^2)}\log \sigma^2\right) -\frac{1}{2}\log |\Psi| \nonumber \\
&-\frac{1}{2}E_{q^*(\sigma^2)}\left(\frac{1}{\sigma^2}\right)E_{q^*(\mathbf{Z}_i)\cdot q^*(\boldsymbol{\beta}_i)}\big[(\mathbf{y}_i - \boldsymbol{g}_i)'\Psi^{-1}(\mathbf{y}_i - \boldsymbol{ g}_i)  \big]\Big\}. \nonumber
\end{align}

Therefore, at each iteration, we update the ELBO using the obtained variational distributions $q^*$ and the current fixed estimate of $w$. The variational EM algorithm stops when the difference between the ELBO values of consecutive iterations becomes negligible or the maximum number of iterations is achieved.

\begin{algorithm}
\caption{The variational EM algorithm for basis function selection for functional data representation with correlated errors}
\label{alg:vem}
\begin{algorithmic}[1]
\State Set hyperparameter values for the prior distributions;
\State Assign initial values to $p^*_{11}, \dots, p^*_{K1}, \dots, p^*_{1m}, \dots, p^*_{Km}, \delta^*_2$, $\lambda^*_2$, and $w$;
\While{$\text{ELBO}^{(c)} - \text{ELBO}^{(c - 1)} > \text{tolerance}$ and $c \leq N_{\text{iter}}$}
  \For{$i = 1, \ldots, m$}
     \State Update the parameters $q^*(\boldsymbol{\beta}_i)$ using \eqref{eq_qbeta_mu} and \eqref{eq_qbeta_sig};
  \EndFor
  \State Update the parameters of $q^*(\sigma^2)$ using \eqref{eq_sigma2_d1} and \eqref{eq_sigma2_d2};
  \State Update the parameters of $q^*(\tau^2)$ using \eqref{eq_tau2_l1} and \eqref{eq_tau2_l2};
    \For{$i = 1, \ldots, m$}
        \For{$k = 1, \ldots, K$}
            \State Update the parameters of $q^*(\theta_{ki})$ using \eqref{eq_theta_a1} and \eqref{eq_theta_a2};
            \State Update the parameter of $q^*(Z_{ki})$ using \eqref{eq:pki}.
        \EndFor
    \EndFor
    \State Update $\text{ELBO}^{(c)}$ using \eqref{eq:elbo_1}-\eqref{eq:elbo_6};
    \State Update $w$ using a quasi-Newton method;
    \State Update $\text{ELBO}^{(c)}$ using the obtained $w$.
  \EndWhile
\end{algorithmic}
\end{algorithm}

\section{Simulations}
\label{sec:sims}
We conduct experiments on synthetic datasets to evaluate the smoothing performance of our proposed method and its ability to select the correct set of basis functions from a set of candidates while accounting for within-curve correlation. In each simulated scenario, curves are generated according to  \eqref{eq:model}. To account for various properties of functions, such as periodicity, we consider two types of functions, $g_i(\cdot)$, in \eqref{eq:model}: one constructed by a linear combination of B-splines and the other by a linear combination of Fourier basis functions. 

For each scenario, we generate 100 datasets, each consisting of $m = 5$ curves, with $n_i = 100$ for all $i$, where the observed values are equally spaced along the curves. To simplify the data generation process, we use the same mean function for all five curves in each simulated dataset, i.e., $g_i(t_{ij}) = g(t_{ij})$ for $i = 1, \dots, 5$. This data generation process leads to replicates sharing the same underlying mean function $g(\cdot)$, allowing us to compute and display the estimated mean function across all five replicates. The results from this approach are presented in Sections \ref{sec:sim12} and \ref{sec:sim3}. However, during estimation, our proposed method smooths each curve individually, with some examples shown in the Supplementary Material.

We set the hyperparameters $\mu_{ki} = 0.5$ for all $k,i$ and $\lambda_1 = \lambda_2 = 10^{-6}$. For initializing our VB algorithm, we set the parameter $\delta_2^{*}$ in the variational distribution of $\sigma^2$, in such way that its mean matches the true noise variance used when generating the datasets, while ensuring that the variance of this distribution is relatively large. For $\lambda_2^*$ and $w$, which are the scale parameter of the variational distribution of $\tau^2$ and the correlation decay parameter, respectively, we set their initial values arbitrarily. Lastly, for the probability $p_{ik}$ that the $k$-th basis is included in the representation of curve $i$, we initially assume that all basis functions should be included, setting $p_{ik} = 1$ for all $i,k$. The algorithm runs with an ELBO convergence threshold of $0.01$ and a maximum of 100 iterations.

To assess our method's performance, we present the mean estimates of the basis function coefficients and their respective standard deviations over the 100 simulated datasets. Specifically, since the curves are generated with same mean functional $g(\cdot)$, the estimates of the basis function coefficients are obtained by averaging the basis coefficients' estimates across the five curves. In particular, we first compute the mode $(\hat{Z}_{ki})$ of the variational distribution of $Z_{ki}$ and the mean ($\mu_{\beta_{ki}}$ as the $k$-th entry in \eqref{eq_qbeta_mu}) of the variational distribution of $\beta_{ki}$ to obtain the estimate of the $k$-th basis coefficient for each curve, defined by $\hat\xi_{ki} = \mu_{\beta_{ki}} \times \hat{Z}_{ki}$. Then, based on the $\hat\xi_{ki}$ values, we compute the estimate of the $k$-th basis coefficient as the average of the $\hat\xi_{ki}$s over the five curves, that is, $\hat\xi_k = \frac{1}{m}\sum_{i = 1}^m\hat\xi_{ki}$. Since we consider 100 simulated datasets, this procedure results in 100 estimates for each basis function coefficient, which we use to compute their mean and standard deviation.

Additionally, based on the estimates of the basis function coefficients, $\hat\xi_{k}$s, obtained as described previously, we present the estimated mean functional and a $95\%$ equal-tailed credible band for one simulated dataset. To construct the equal-tailed credible band, we adapt the approach used in \cite{sousa2024} to consider the variational distributions. Specifically, we sample 200 values from the variational distributions of each $\mathbf{Z}_{i}$ and $\boldsymbol{\beta}_i$. Using these samples, we compute the estimate of each basis coefficient, $\hat\xi_k$. Then, based on these estimates we generate 200 curves evaluated at the same points. For each evaluation point, we compute the $2.5\%$ and $97.5\%$ quantiles of the function values to form the credible band.

In Sections \ref{sec:sim12} and \ref{sec:sim3}, we provide in details the data generation procedure used in the simulated scenarios and comment on the performance of our model in performing basis selection, specifically its ability in identifying the correct set of basis functions from a set of candidates while accounting for within-curve correlation.

\subsection{Simulated Scenarios 1 and 2}
\label{sec:sim12}
In Scenarios 1 and 2, we generate the curves similar to \cite{sousa2024}, but we consider a within-curve correlation with a correlation decay parameter of $w = 6$ in $\psi(\cdot, \cdot)$. More specifically, the curves are generated by a linear combination of cubic B-splines with known true coefficients, plus some random noise generated by a Gaussian process as described in Section \ref{sec:model}. For Scenario 1, we consider $\sigma^2 = 0.1^2$, while for Scenario 2, we assume $\sigma^2 = 0.2^2$.

To assess the performance of our proposed method in selecting the correct number and the true set of basis functions, we deliberately set four basis coefficients to zero. This deliberate manipulation allows us to evaluate how well the method identifies and recovers the true underlying structure of the data. 

The curves are generated at each evaluation point $t_{ij}$ as follows:
\begin{equation*}
y(t_{ij}) = (-2, 0, 1.5, 1.5, 0, -1, -0.5, -1, 0, 0)'\mathbf{B}(t_{ij}) + \varepsilon(t_{ij}) ,
\end{equation*}
where $\varepsilon(t_{ij}) \sim GP(0, \sigma^2\psi(\cdot,\cdot))$ and $\mathbf{B}(t_{ij}) = (B_1(t_{ij}), \dots, B_K(t_{ij}))\trp$ is the vector of the K known B-splines evaluated at $t_{ij}$. Furthermore, without loss of generality, we assume that the curves are observed in the interval $[0, 1]$ and at the same evaluation points, that is, $t_{ij} = t_j$ for all $i = 1, \dots, 5$

Table \ref{tab:sim12} shows the mean estimated coefficients for the basis functions computed as described in Section \ref{sec:sims}, along with their respective standard deviations (SD). Notably, the estimated coefficients closely align with the true values in both scenarios with higher standard deviations for scenario 2 due to the higher noise variance.

\begin{table}
\centering
\caption{Simulated Scenarios 1 and 2. Mean estimates for the basis function coefficients (Mean) over the 100 simulated datasets accordingly to the data dispersion and when $w = 6$. The mean estimates are computed as $\frac{1}{100}\sum_{s = 1}^{100} \hat\xi_{k}^s$, where $\hat\xi_{k}^s = \frac{1}{5}\sum_{i = 1}^5 \hat\xi_{ki}^s$ as described in Section \ref{sec:sims} and $s$ indicates the $s$-th simulated dataset. We also present the respective standard deviations (SD). For comparison, the true basis coefficients are included as well.} 
\label{tab:sim12}
\begin{tabular}{lrrlrl}
  \toprule
 &  & \multicolumn{2}{c}{$\sigma = 0.1$} & \multicolumn{2}{c}{$\sigma = 0.2$} \\
 \cmidrule(r){3-4}\cmidrule(l){5-6}
 & True  & Mean & \multicolumn{1}{l}{SD} & Mean & \multicolumn{1}{l}{SD}  \\ 
  \midrule
$\hat\xi_1$    & $-2.0 $ & $ -1.9940$ &0.0400 &$-1.9753$ &$0.0779$ \\
$\hat\xi_2$    & $ 0.0 $ & $  0.0060$ &0.0484 &$ 0.0189$ &$0.0958$ \\
$\hat\xi_3$    & $ 1.5 $ & $  1.4727$ &0.0654 &$ 1.4261$ &$0.1272$ \\
$\hat\xi_4$    & $ 1.5 $ & $  1.4961$ &0.0606 &$ 1.4735$ &$0.1177$ \\
$\hat\xi_5$    & $ 0.0 $ & $ -0.0004$ &0.0430 &$ 0.0026$ &$0.0824$ \\
$\hat\xi_6$    & $-1.0 $ & $ -0.9785$ &0.0586 &$-0.9432$ &$0.1161$ \\
$\hat\xi_7$    & $-0.5 $ & $ -0.4966$ &0.0549 &$-0.4411$ &$0.1417$ \\
$\hat\xi_8$    & $-1.0 $ & $ -0.9774$ &0.0610 &$-0.9352$ &$0.1266$ \\
$\hat\xi_9$    & $ 0.0 $ & $  0.0029$ &0.0526 &$ 0.0122$ &$0.1026$ \\
$\hat\xi_{10}$ & $ 0.0 $ & $  0.0024$ &0.0396 &$ 0.0102$ &$0.0786$\\
   \bottomrule
\end{tabular}
\end{table}

In Figures 1a and 1c, we present the estimated mean curve over the $m = 5$ replicates, $\hat{g}$, along with a $95\%$ equal-tailed credible band for one simulated dataset under both scenarios ($\sigma = 0.1$ and $\sigma = 0.2$) obtained from our proposed variational EM method which accounts for the data correlation structure. Notably, the estimated mean curve closely aligned with the true curve consistently across the evaluation points. Regarding the credible band, we observe that the band is relatively narrow, and the band widens with a higher noise variance. In particular, for this scenario, Figures 1-3 in the Supplementary Material show the estimated smooth curve corresponding to each individual replicate and each individual mean across all simulated datasets, which also aligns to the true curve.

\begin{figure}
    \centering
    \begin{subfigure}[b]{.49\textwidth}
        \centering
        \includegraphics[width = \textwidth]{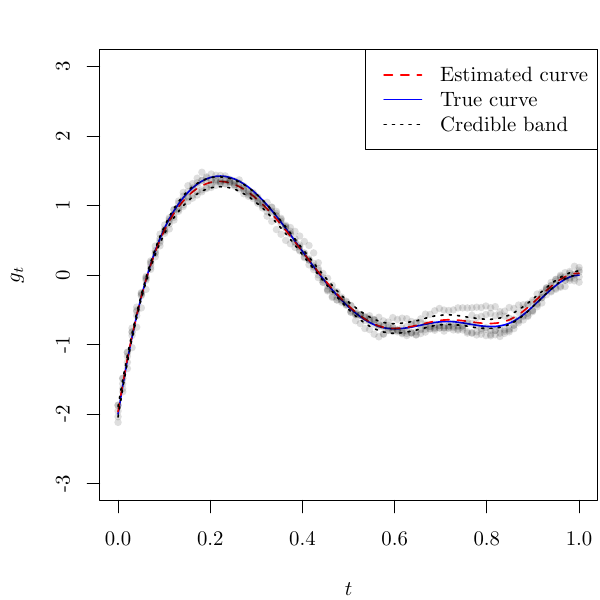}
        \caption{Correct correlation structure ($\sigma = 0.1$)}
    \end{subfigure}
     \begin{subfigure}[b]{.49\textwidth}
        \centering
        \includegraphics[width = \textwidth]{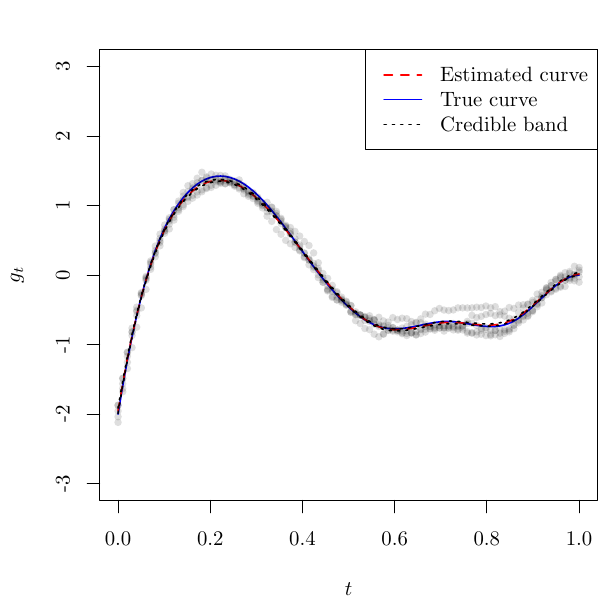}
        \caption{Misspecified correlation structure ($\sigma = 0.1$)}
    \end{subfigure}
    \begin{subfigure}[b]{.49\textwidth}
        \centering
        \includegraphics[width = \textwidth]{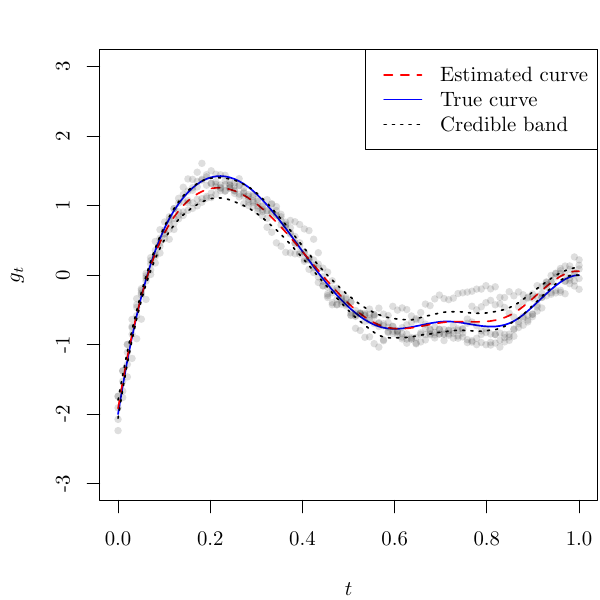}
        \caption{Correct correlation structure ($\sigma = 0.2$)}
    \end{subfigure}
    \begin{subfigure}[b]{.49\textwidth}
        \centering
        \includegraphics[width = \textwidth]{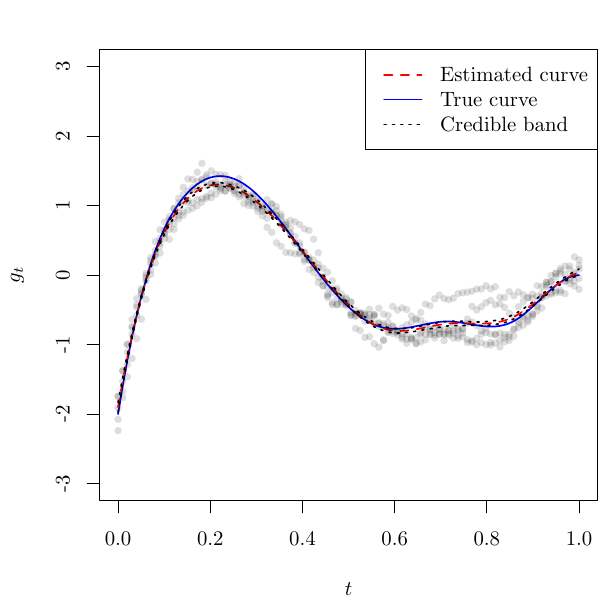}
        \caption{Misspecified correlation structure ($\sigma = 0.2$)}
    \end{subfigure}
    \caption{Simulated Scenarios 1 and 2. True (blue) and estimated (red) mean curves with a 95\% equal-tailed credible band (dashed lines) for one simulated dataset accordingly to the data dispersion, $K = 10$ and $w = 6$. The gray points correspond to the generated points from the five curves. Figures (a) and (c) refer to our proposed method that estimates the data correlation structure for $\sigma = 0.1$ and $\sigma = 0.2$ respectively. Figures (b) and (d) were obtained by running VB algorithm assuming independence among observations within a curve for $\sigma = 0.1$ and $\sigma = 0.2$ respectively.}
\label{fig:curve_sim12}
\end{figure}

Furthermore, Figure \ref{fig:curve_sim12} also compares the results when accounting for the data correlation structure using our proposed method with those obtained when the covariance structure is misspecified by assuming the observations within a curve are uncorrelated. When the correlation in the data is misspecified, the uncertainty in the results is underestimated, leading to narrower credible bands, as shown in Figures 1b and 1d. To further investigate this behaviour we observe that the mean estimated noise variances for the proposed method are $0.0097$ and $0.0388$ for the scenarios with $\sigma^2 = 0.1^2$ and $\sigma^2 = 0.2^2$, respectively across the 100 simulated datasets. In contrast, for the misspecified scenarios, the mean estimated noise variances across the simulated datasets are $0.0044$ and $0.0178$ for $\sigma^2 = 0.1^2$ and $\sigma^2 = 0.2^2$, respectively. This demonstrates that the noise variance is underestimated in the misspecified scenarios. This behaviour was also observed when considering a Gibbs sampler (results not shown) under a misspecified correlation structure in the estimation process.

\begin{figure}
    \centering
    \begin{subfigure}[b]{.49\textwidth}
        \centering
        \includegraphics[width = \textwidth]{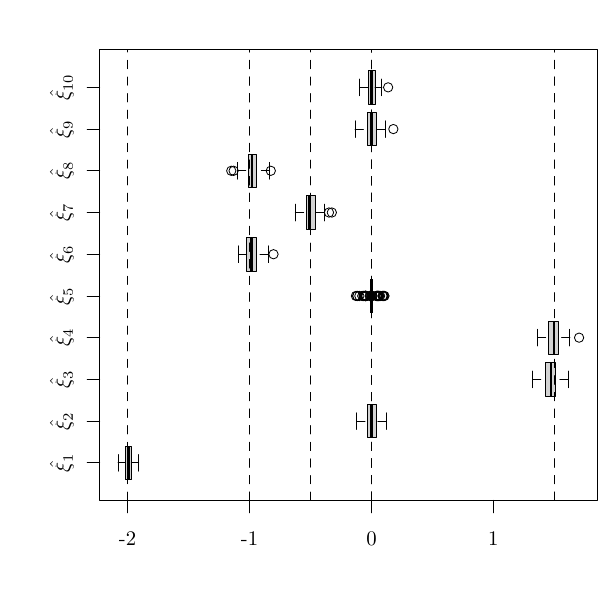}
        \caption{$\sigma = 0.1$}
    \end{subfigure}
    \begin{subfigure}[b]{.49\textwidth}
        \centering
        \includegraphics[width = \textwidth]{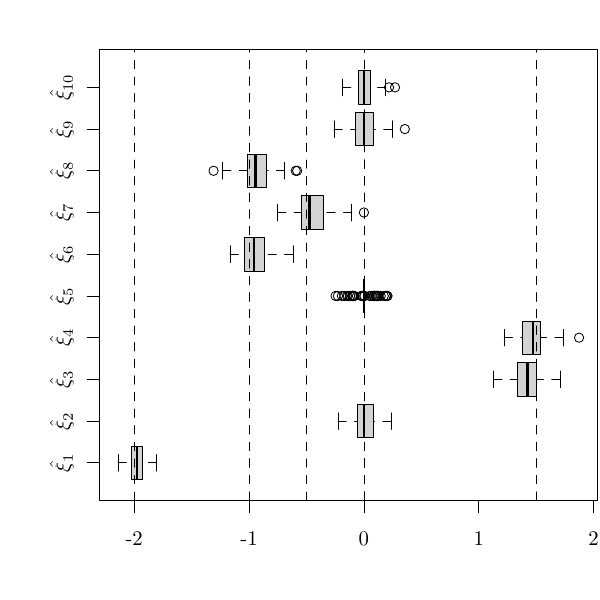}
        \caption{$\sigma = 0.2$}
    \end{subfigure}
    \caption{Simulated Scenarios 1 and 2. Boxplots of the estimates of the basis function coefficients, $\hat{\xi}_ks$, generated from 100 simulated datasets accordingly to the data dispersion, $K = 10$ and $w = 6$. The dashed lines correspond to the true values of the basis coefficients.}
    \label{fig:boxplot_sim12}
\end{figure}

Additionally, our model effectively selects the correct set of basis functions, as demonstrated in Figure \ref{fig:boxplot_sim12}. In this figure, boxplots with values distant from zero correspond to true non-zero basis coefficients used to generate the data. We also evaluate the true and false positive rates for the basis functions selected via sensitivity, specificity and accuracy. The sensitivity (true positive rate) refers to the proportion of basis functions correctly identified as being used in the data generation of the mean curve. The specificity (1 - false positive rate), where the false positive rate represents the proportion of basis functions incorrectly selected when they should not have been. In Table \ref{tab:sims_acc}, we display the average sensitivity, specificity, and accuracy taken over the 100 simulated datasets for the selected basis functions used to estimate the mean curve. As expected, these measures exhibit a minor decrease in the presence of a higher dispersion. However, we observe high precision in recovering the true set of basis functions regardless of the data dispersion. Finally, the estimated correlation decay parameter obtained using our proposed method has a median of $6.1553$ (IQR = 0.5006) and $6.1414$ (IQR = 0.4952) across the 100 simulated datasets for scenarios with $\sigma^2 = 0.1^2$ and $\sigma^2 = 0.2^2$ respectively.

\begin{table}
\centering
\caption{Simulated Scenarios 1, 2 and 3. Sensitivity (Sens), specificity (Spec) and accuracy (Acc) taken as the average over the 100 simulated datasets for the selected basis functions used in the estimation of the mean curve accordingly to the data dispersion. Scenarios 1 and 2 considered B-splines basis functions (Section \ref{sec:sim12}) and Scenario 3 considered Fourier basis functions (Section \ref{sec:sim3}). For all scenarios $w = 6$ was used for data generation.} 
\label{tab:sims_acc}
\begin{tabular}{l|lll}
  \toprule
   Scenario &  Sens & Spec & Acc\\
   \midrule
1 (B-splines, $\sigma = 0.1$) & 1.0000 & 0.9250 & 0.9700 \\
2 (B-splines, $\sigma = 0.2$) & 0.9750 & 0.8975 & 0.9400 \\
3 (Fourier, $\sigma = 0.1$)   & 1.0000 & 0.9950 & 0.9960 \\
   \bottomrule
\end{tabular}
\end{table}

\subsection{Simulated Scenario 3}
\label{sec:sim3}
For simulation Scenario 3, we evaluate our method's performance when we have periodic data. We adapt the second simulated study from \cite{sousa2024} to include within-curve correlation by assuming a correlation decay parameter $w = 6$ in $\psi(\cdot, \cdot)$. The data is generated as a linear combination of trigonometric functions with a noise variance of $\sigma^2 = 0.1^2$ and same evaluation points $t_{ij} = t_j \in [0, 2\pi]$ for all $i = 1, \dots, 5$.

The curves are generated as follows:
\begin{equation*}
y(t_{ij}) = \cos(t_{ij}) + \sin(2t_{ij}) + \varepsilon(t_{ij}) ,
\end{equation*}
with $\varepsilon(t_{ij}) \sim GP(0, \sigma^2\psi(\cdot,\cdot))$. 

\begin{figure}[ht]
    \centering
    \begin{subfigure}[b]{.49\textwidth}
        \centering
       \includegraphics[width = \textwidth]{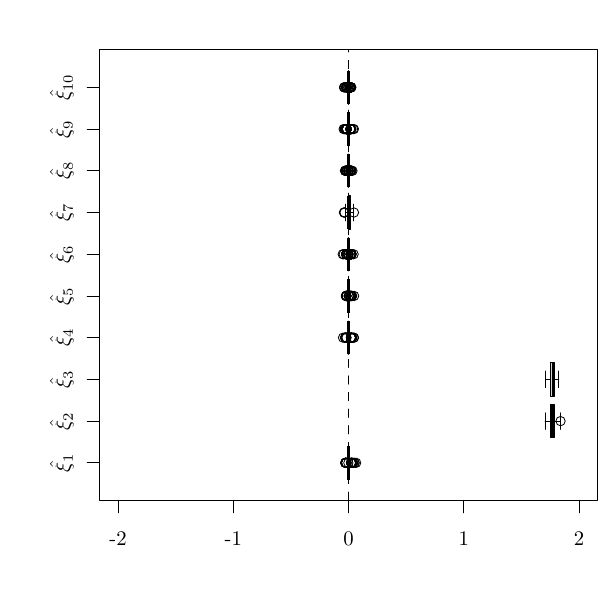}
        \caption{}
        \label{fig:boxplot_sim3}
    \end{subfigure}
    \begin{subfigure}[b]{.49\textwidth}
        \centering
        \includegraphics[width = \textwidth]{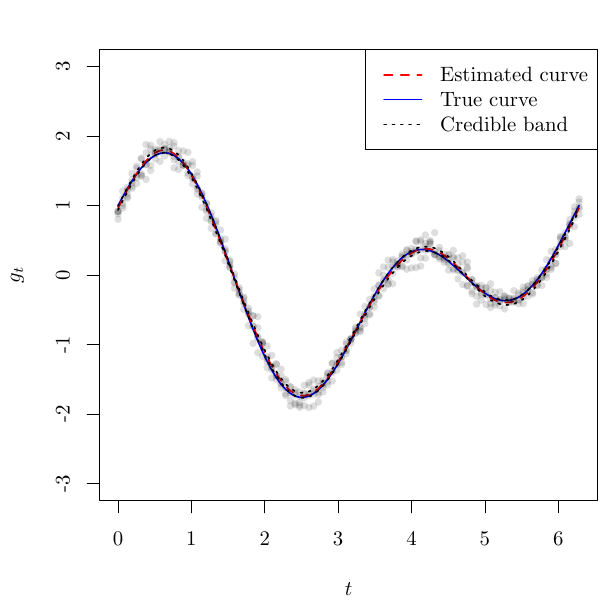}
        \caption{}
        \label{fig:curve_sim3}
    \end{subfigure}
    \caption{Simulated Scenario 3. Figure (a) shows the boxplots of basis coefficients estimates, $\hat{\xi}_ks$, generated from 100 simulated datasets when $\sigma = 0.1$, $K = 10$, and $w = 6$. The dashed line at zero shows which basis functions are being removed when representing the functional data. The gray points correspond to the generated points from the five curves. Figure (b) displays true (blue) and estimated (red) mean curves with a 95\% equal-tailed credible band (dashed lines) for one simulated dataset when $\sigma = 0.1$, $K = 10$, and $w = 6$.}
\end{figure}

For periodic data, Fourier basis functions are typically used to represent the data. Given the simplicity of the generated curves, we employ ten Fourier basis functions. Similar to previous scenarios, Table \ref{tab:sim3} presents the mean estimated basis coefficients as the simple average across the five curves and simulated datasets. Figure  $\ref{fig:boxplot_sim3}$ shows the boxplots of $\hat{\xi}_{k}$ across all simulated datasets. Notably, only two basis functions showed non-zero values, indicating that these basis functions sufficiently represent the data. Table \ref{tab:sim3} also includes the standard deviations of the average estimated basis coefficients, demonstrating high precision in the estimates. We also observe a high sensitivity, specificity and accuracy as shown in Table \ref{tab:sims_acc}, indicating a high accuracy in recovering the correct set of basis functions used in the mean curve for the data generation.

Figure \ref{fig:curve_sim3} shows the estimated mean curve from one simulated dataset, calculated as the average of the results obtained from the five curves along with a $95\%$ equal-tailed credible band. The estimated mean curve closely align with the true one, with a narrow credible band. Additionally, our method satisfactorily estimated the correlation decay parameter, showing a median of $6.4890$ (IQR = 0.7818) across the 100 simulated datasets.

\begin{table}
\centering
\caption{Simulated Scenario 3. Mean estimates for the basis function coefficients (Mean) over the 100 simulated datasets when $\sigma = 0.1$, $K = 10$, and $w = 6$. The mean estimates are computed as $\frac{1}{100}\sum_{s = 1}^{100} \hat\xi_{k}^s$, where $\hat\xi_{k}^s = \frac{1}{m}\sum_{i = 1}^M \hat\xi_{ki}^s$ as described in Section \ref{sec:sims}. We also display their respective standard deviations (SD).}
\label{tab:sim3}
\begin{tabular}{lcccccccccc}
  \toprule
       & $\hat\xi_1$ & $\hat\xi_2$ & $\hat\xi_3$ & $\hat\xi_4$ & $\hat\xi_5$ & $\hat\xi_6$ & $\hat\xi_7$ & $\hat\xi_8$ & $\hat\xi_9$ & $\hat\xi_{10}$ \\
  \midrule
  Mean & 0.0026 & 1.7678 & 1.7700 & 0.0008 & 0.0016 & 0.0009 & 0.0037 & -0.0008 & -0.0001 & 0.0011 \\
  SD   & 0.0169 & 0.0243 & 0.0242 & 0.0138 & 0.0123 & 0.0137 & 0.0170 & 0.0132 & 0.0150 & 0.0123 \\
  \bottomrule
\end{tabular}
\end{table}

\section{Real Data Analysis}

\subsection{Motorcycle dataset}
\label{sec:mcycle}

First, we analyze the \pck{mcycle} dataset available  in the \pck{faraway} \R package, which has been extensively analyzed in the literature \citep{silverman1985, eilers1996, storlie2010, schiegg2012, mraoui2024, goepp2025}. A detailed description of the data is provided by \cite{souza2014}. The dataset consists of 133 measurements of head acceleration (in units of g) taken over time (in milliseconds) during a simulated motorcycle accident. We jittered the time points by adding a small random noise to them, because some time points have multiple acceleration measurements, which impedes the use of smoothing methods. 

We applied our proposed method to represent the motorcycle dataset as a smooth function considering three initial sets of cubic B-splines basis functions to perform selection. These correspond to $K = 15, 20, 30$ equally spaced  basis functions. We select the optimal set based on the generalized cross-validation (GCV) criterion \citep{craven1978}. The GCV values for $K = 15, 20, 30$ were $520.0588$, $518.5663$, and $523.3633$ respectively. Therefore, we chose the results for the set with $K = 20$ as the optimal set to represent the data. Additionally, to initialize the scale parameter $\delta_2^*$ of the variational distribution of the noise variance $\sigma^2$ in our variational EM algorithm, we consider the mean square error from a regression splines with 50 basis functions. Specifically, $\delta_2^*$ is set so that the initial mean of the variational distribution of $\sigma^2$ matches the mean square error obtained from the fitted regression splines. We set the correlation decay parameter $w = 10$ and $\lambda_2^* = 100$. For the inverse-gamma prior distribution of $\sigma^2$, we consider a large prior mean (e.g., 50) and a large prior variance (e.g., 300).  For the inverse-gamma prior distribution of $\tau^2$, we consider a non-informative prior. As in the simulated studies, we initially include all basis functions in the data representation. The ELBO convergence threshold was set to 0.001.

We compare our proposed method to four commonly used techniques for smoothing functional data. For all methods, we employ the same initial set of $K = 20$ basis functions. For the Bayesian LASSO \citep{park2008} and LASSO \citep{tibshirani1996}, the basis function matrix serves as the input for the design matrix. Furthermore, for LASSO and smoothing splines, we consider a grid of values to determine the smoothing parameter. The model based on the parameter that yielded the best performance was then used for comparison with the other methods.

The methods are compared based on the adjusted $R^2$ and by visually inspecting the smoothed data. Overall, our proposed method exhibited the highest adjusted $R^2 = 0.7860$, followed closely by smoothing splines ($R^2 = 0.7787$), LASSO ($R^2 = 0.7698$), regression splines ($R^2 = 0.7698$) and Bayesian LASSO ($R^2 = 0.7452$). Although the methods showed similar performances based on adjusted $R^2$ values, our method offers advantages over the other techniques, such as parsimony and interpretability. Moreover, our method was the only one that reduced the number of basis functions used to fit the data, using only five out of the initial set of 20 basis functions. In contrast, smoothing splines that only shrink the coefficients does not perform basis selection. Therefore, utilizing all 20 basis functions. Bayesian LASSO and LASSO, which also incorporate regularization and can perform basis function selection, end it up utilizing all 20 basis functions.

It is important to note that when the number of variables is large but only a small subset is truly relevant, as in this the case with the motorcycle dataset, the application of LASSO for variable selection may present limitations. Specifically, LASSO may either fail to effectively identify the truly relevant subset of variables, as observed in its application to the motorcycle data, or excessively shrink the coefficients of the variables it retains in order to sufficiently
reduce the set of chosen variables \citep{hastie2015}.

Additionally, our method was the only one to capture the expected behavior present in the data. Specifically, since the data measures head acceleration in a simulated motorcycle accident, it is expected that head acceleration should be zero at the start of the experiment, as the impact has not yet occurred. Similarly, after the impact, head acceleration should return to zero. Our method was the only one to capture this expected behavior.

\begin{figure}[ht]
    \centering
    \includegraphics[width = 0.8\textwidth]{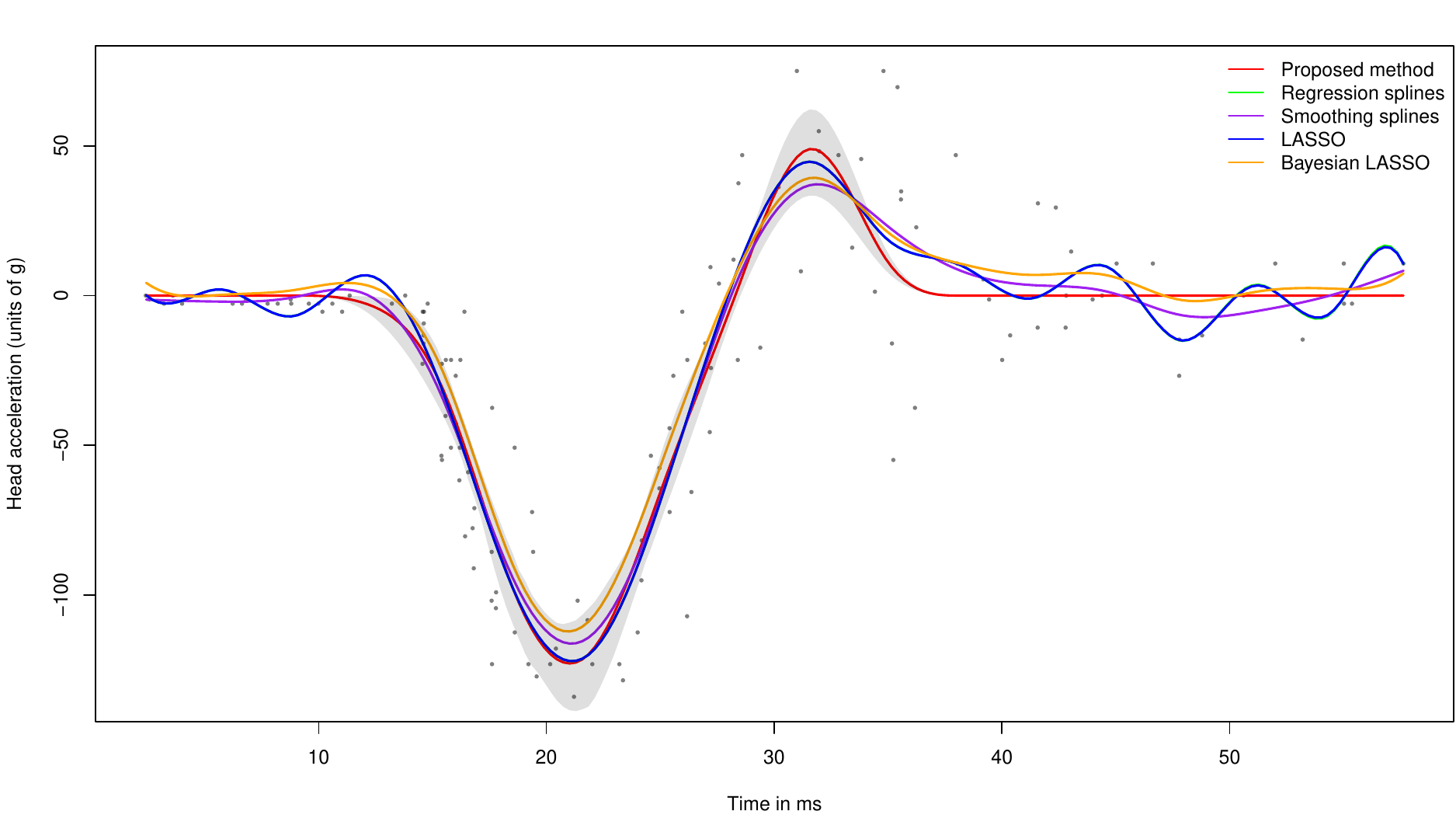}
    
    \includegraphics[width = 0.8\textwidth]{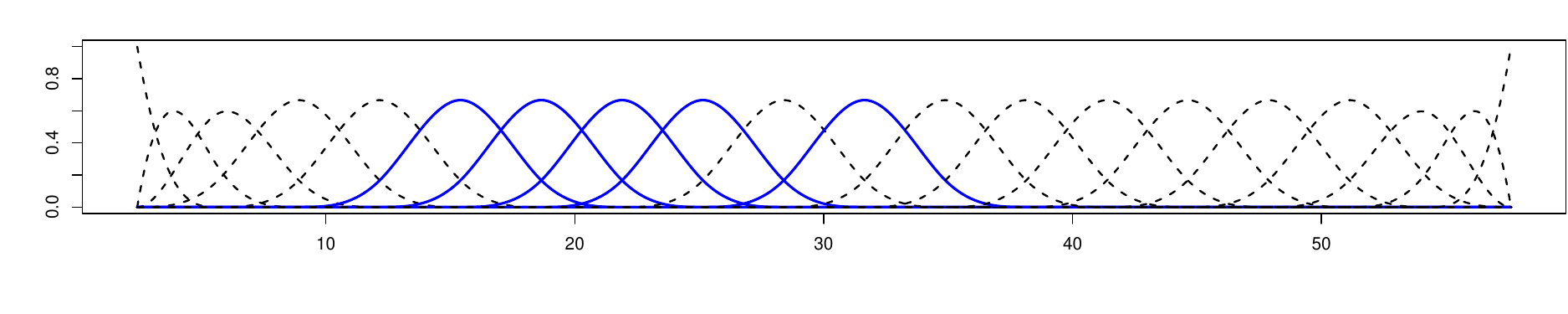}

    \caption{The $x$-axis represents time in milliseconds (ms), while the $y$-axis shows head acceleration in units of g. The estimated mean curves are plotted using our proposed method (red), regression splines (green), smoothing splines (purple), LASSO (blue), and Bayesian LASSO (orange). A 95\% credible band (shaded area) is also provided, based on our proposed method, for the motorcycle dataset. The basis functions selected by our method are highlighted as the solid blue curves, whereas the removed basis functions are shown as dashed black curves.}
    \label{fig:data}
\end{figure}

Figure \ref{fig:data} shows the mean curves estimated by all methods with a $95\%$ credible band for our proposed method (shaded area). Visually, the estimated curves obtained closely align in some regions. However, we note that the Bayesian LASSO tends to penalize certain regions more heavily than the other methods. For example, in the 18-23 ms range, the Bayesian LASSO estimates higher head acceleration than the other methods. Moreover, Figure \ref{fig:data} presents the set of basis functions selected by our method, which are predominantly concentrated in the region where the head acceleration diverges from zero. This suggests that more basis functions are necessary to accurately capture the data patterns in the region.

Despite the estimated within-curve correlation function being negligible ($\hat w = 10^6$), indicating minimal or no correlation in the head measurements, our proposed method remains highly effective. It demonstrates comparable performance to other regularization techniques. This highlights how adaptable and reliable our method is, making it a compelling choice even when within-curve correlation is negligible.

\subsection{Canadian weather temperature dataset}
\label{sec:weather}

Now, considering the \pck{CanadianWeather} dataset available in the \pck{fda} \R package, which comprises daily temperature and precipitation information from 35 different locations in Canada averaged over the period 1960 to 1994, we analyze the daily temperature over 365 days for six specific locations: Montreal, Quebec City, Arvida, Bagottville, Sherbrooke, and Vancouver. Before performing the analysis, the temperature dataset for each of those six locations was standardized with respect to its standard deviation. For the purpose of visualizing the estimated mean curves, we converted the data back to their original scales.

Similar to the analysis of the motorcycle dataset, we applied our method to the temperature dataset considering five initial sets of cubic B-splines basis functions to perform selection. These correspond to $K = 10, 15, 20, 30$, and $50$ equally spaced  basis functions. We select the optimal set based on the generalized cross-validation (GCV) criterion. The initialization of the scale parameter $\delta_2^*$ was also similar to the motorcycle dataset, where we set this parameter, such that the mean of the variational distribution of the noise variance $\sigma^2$ was equal to the mean square error from a regression splines with 50 cubic B-splines basis functions. For the prior of $\sigma^2$, since the data was standardized, we set $\delta_1 = 10$ and  $\delta_2 = 0.09$, resulting in a moderate prior mean. The other parameters and hyperparameters were initialized as in the analysis of the motorcycle dataset.

We observe that the optimal initial set of basis functions is with $K = 30$ for Arvida, Bagottville, Sherbrooke, and Vancouver, while for the other cities, the set with $K = 20$ basis functions were considered to be optimal (see Table 1 in the Supplementary Material). Figure \ref{fig:data_temp_raw} shows the raw and estimated mean curves colored by city, as obtained using our method. Our method effectively captures the temperature trends across all evaluated cities, despite the estimation being performed simultaneously for all cities. Additionally, in Figure \ref{fig:data_temp_2cities_VEM} we display the estimated mean curve for two cities accompanied by a 95\% credible band. The estimated mean curves captures the overall trend in the daily temperature throughout the year, with reasonably narrow credible bands.

\begin{figure}
    \centering
    \includegraphics[width = 0.8\textwidth]{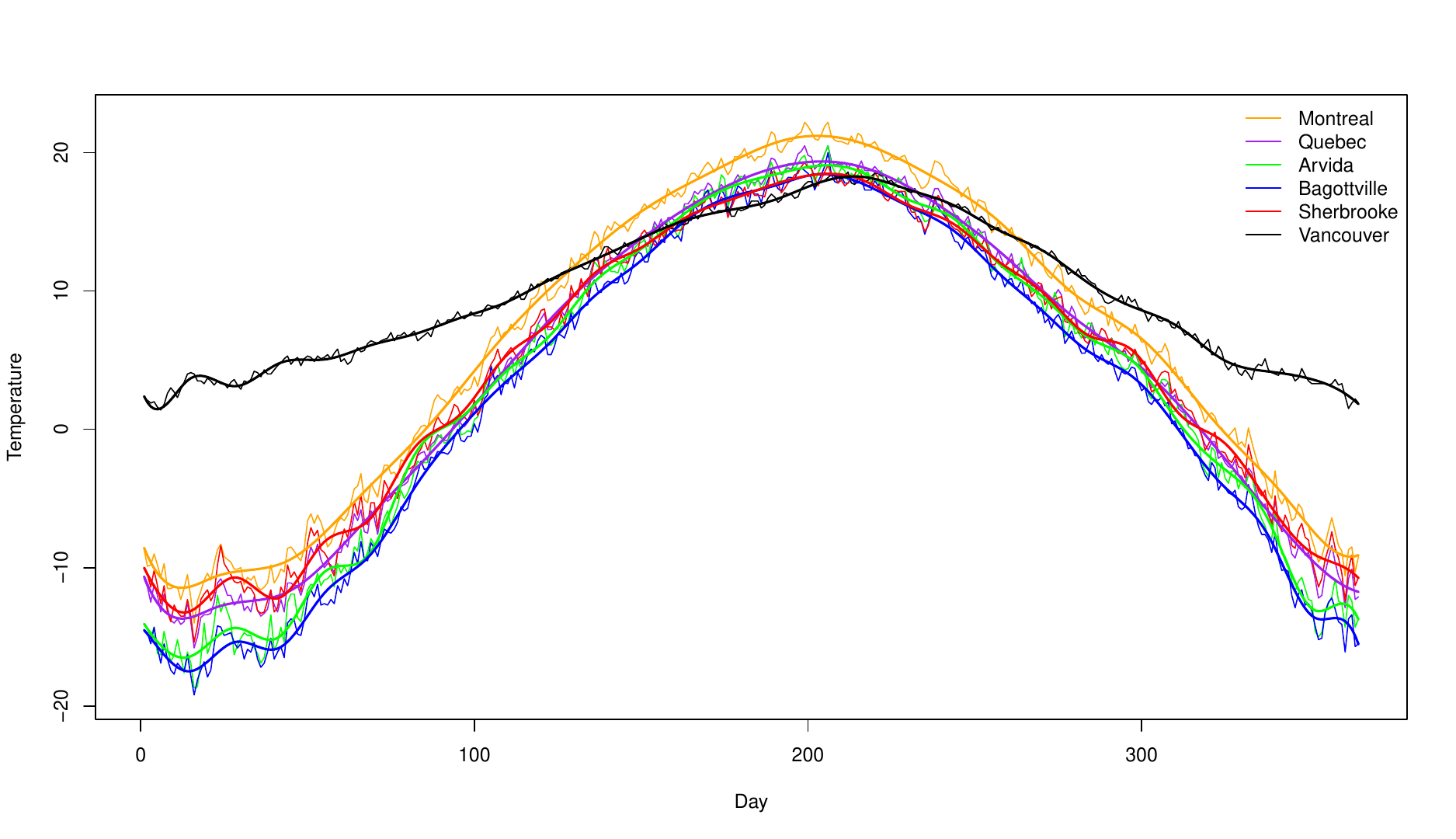}

    \caption{Temperature analysis. The $y$-axis represents temperature in degrees Celsius, while the $x$-axis represents the day of the year. The graph displays the raw and estimated mean curves for daily temperatures throughout the entire year in six cities, with the smoothed lines showing the results of our method.}
    \label{fig:data_temp_raw}
\end{figure}

\begin{figure}
    \centering
    \includegraphics[width = 0.8\textwidth, page = 2]{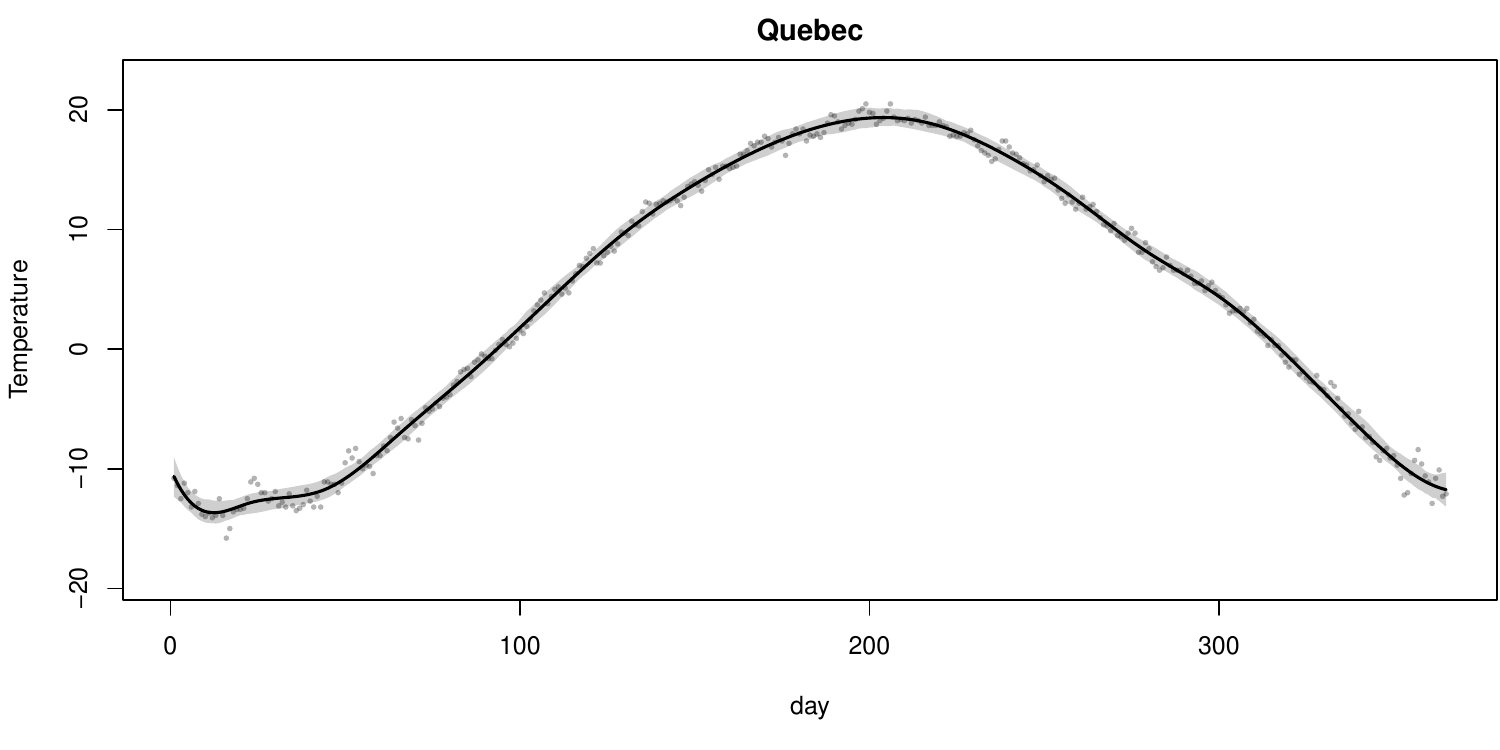}
    \includegraphics[width = 0.8\textwidth, page = 6]{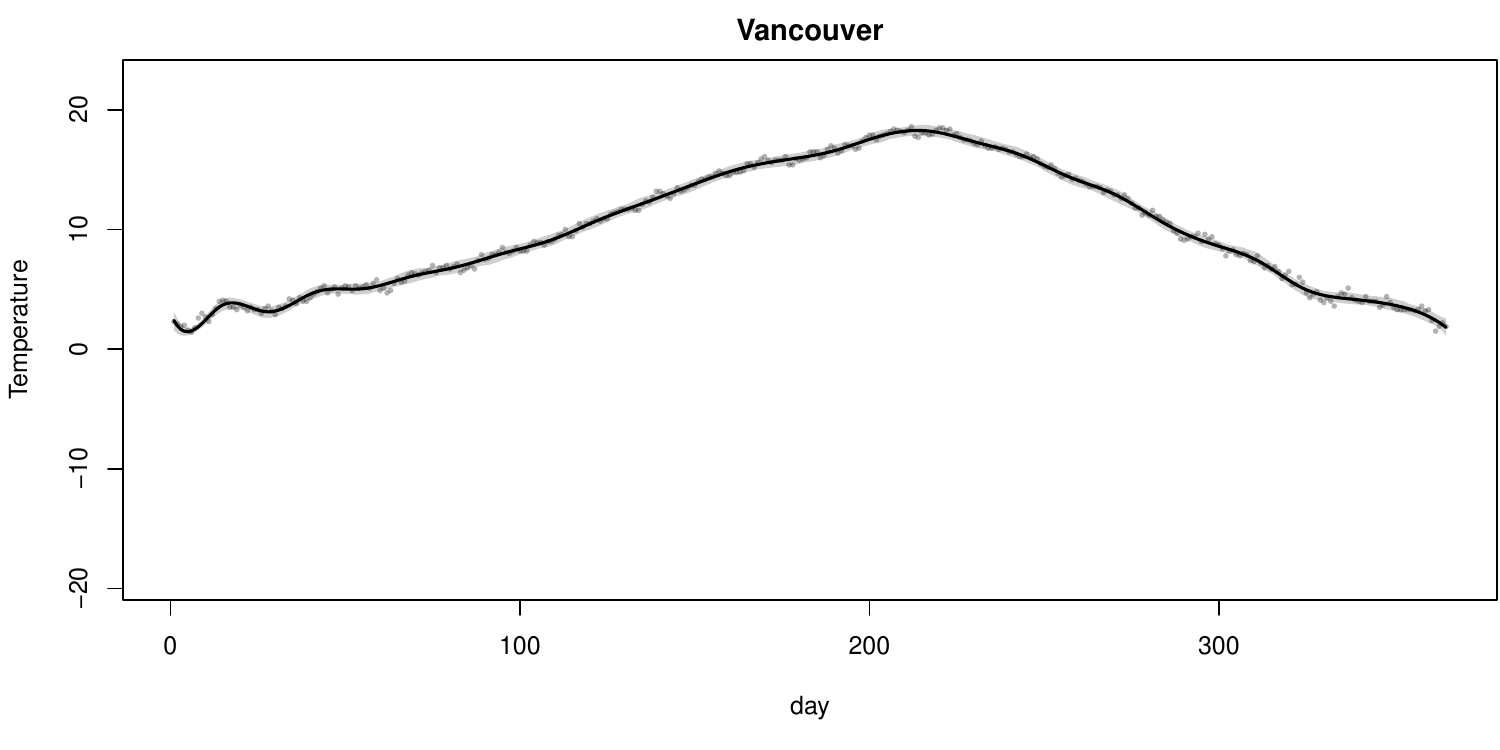}
    \caption{Temperature analysis. The $y$-axis represents temperature in degrees Celsius, while the $x$-axis represents the day of the year. The graph displays the estimated mean curves for the daily temperature throughout the entire year for Quebec City and Vancouver using our method with a $95\%$ credible band (shaded area).}
    \label{fig:data_temp_2cities_VEM}
\end{figure}

To compare our method with regression splines, smoothing splines, Bayesian LASSO, and LASSO, we used the same matrix of basis functions, determined by the optimal number of basis functions for each city across all methods. Our approach offers the advantage of fitting multiple curves simultaneously, while the other techniques require the user to modify the functions available in \R to fit all curves at once. For our analysis, we obtained results for regression splines, smoothing splines, LASSO, and Bayesian LASSO by applying the \pck{bs}, \pck{smooth.basis}, \pck{glmnet} and \pck{blasso} functions from the \R packages \pck{splines}, \pck{fda}, \pck{glmnet}, and \pck{monomvn} respectively, to each curve individually. 

\begin{figure}
    \centering
    \includegraphics[width = 0.8\textwidth, page = 4]{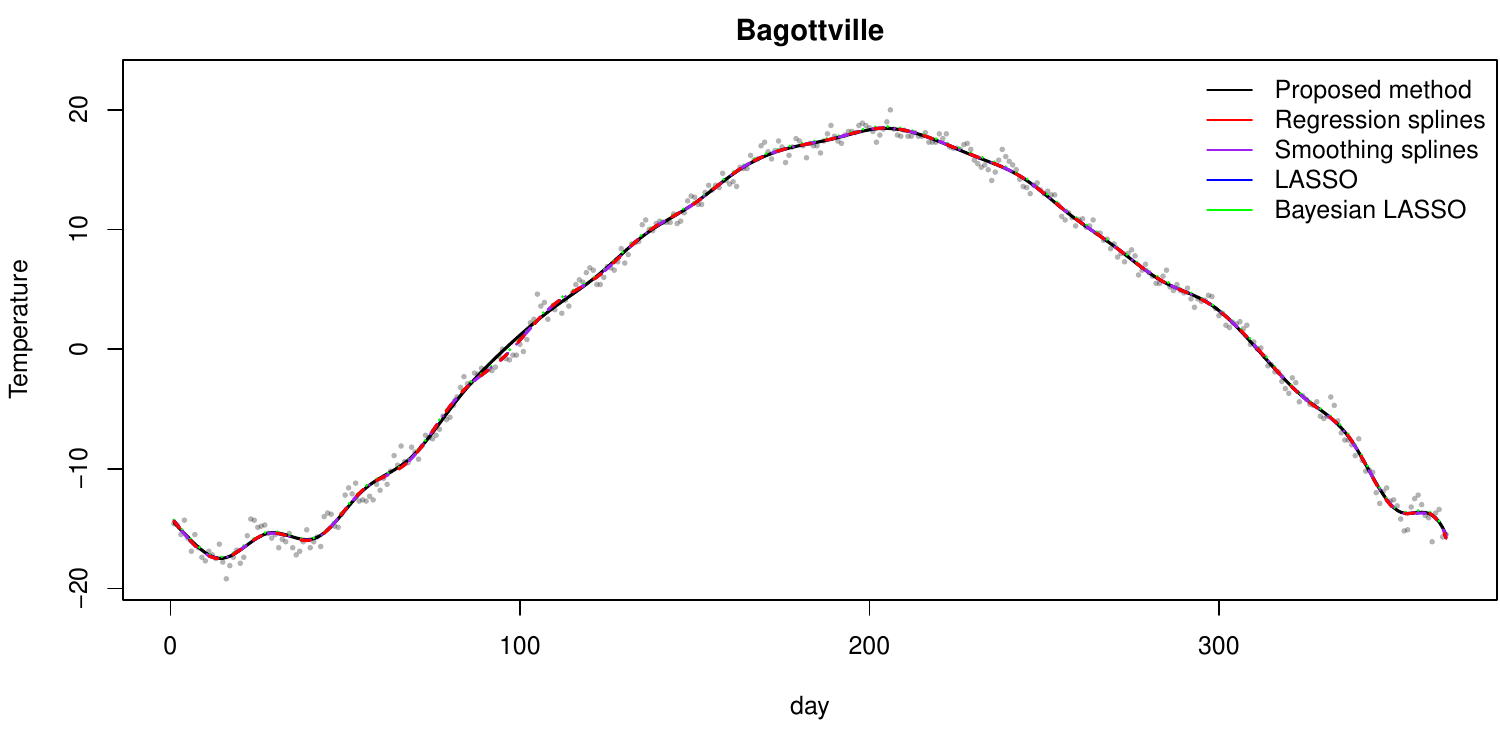}
    \includegraphics[width = 0.8\textwidth, page = 5]{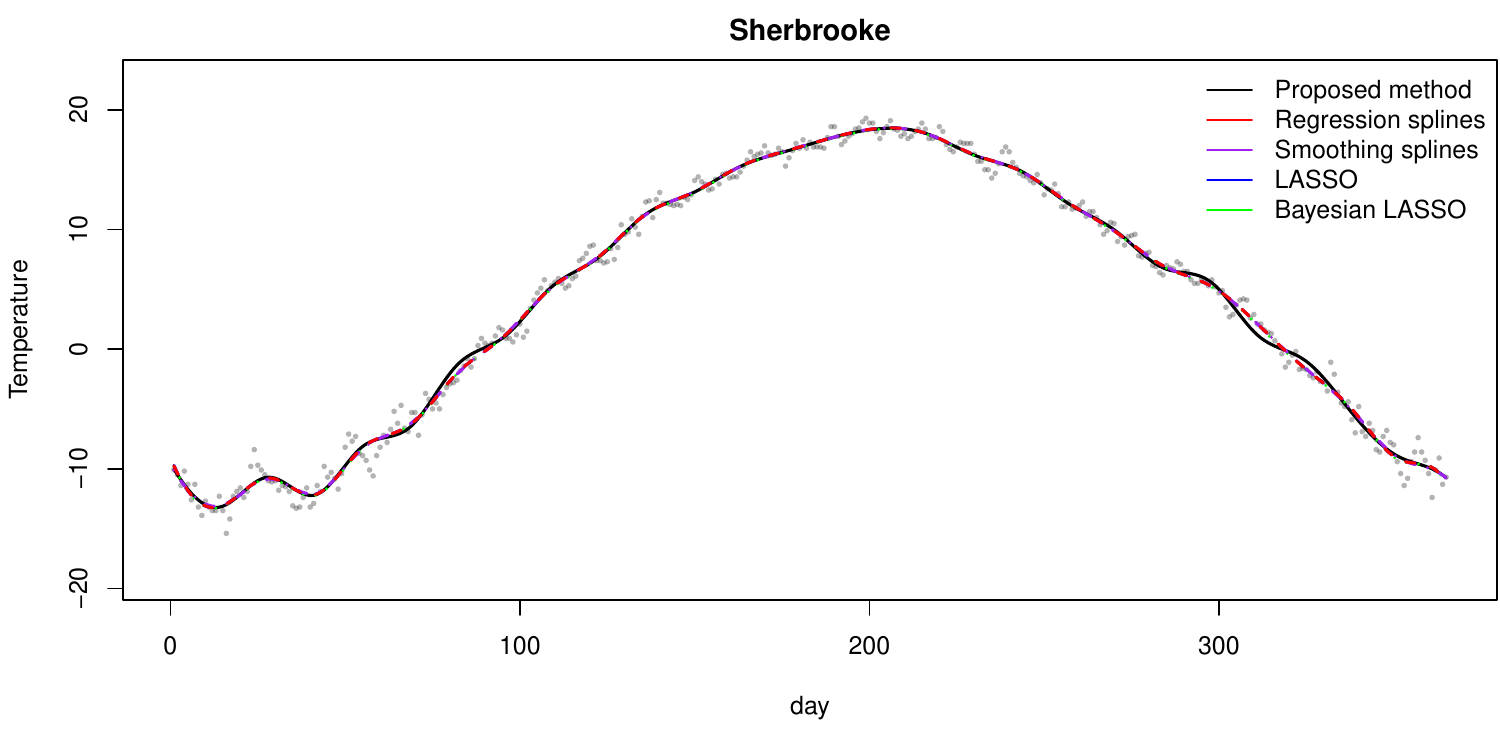}
    \caption{Temperature analysis. Temperature in degrees Celsius in the $y$-axis and day of the year in the $x$-axis. Estimated mean curves for the daily temperature throughout the entire year for Bagottville and Sherbrooke using our proposed method (black), regression splines (red), smoothing splines (purple), LASSO (blue), and Bayesian LASSO (green).}
    \label{fig:data_temp_2cities2}
\end{figure}

As in the motorcycle dataset analysis (Section \ref{sec:mcycle}), our method once again demonstrate competitive performance based on adjusted $R^2$ compared to the other approaches (as shown in Table 2 in the Supplementary Material).
Figure \ref{fig:data_temp_2cities2} and Supplementary Figures 4 and 5 display the estimated mean curves for four cities using each method over the entire year. Overall, the estimated mean curves are closely aligned across all methods, with no significant differences. For Arvida (Supplementary Figure 4) and Sherbrooke (Figure \ref{fig:data_temp_2cities2}), we observe a deviation in the curve obtained from our proposed method around days 100 and 300. This deviation is due to a basis function being removed in those regions, causing the curves to shift closer to zero. However, this result was not observed for the other cities. Regarding the within-curve correlation function, our method estimated the correlation decay parameter $\hat{w} = 161.46$ for when $K = 20$ and $\hat{w} = 152.19$ for $K = 30$, resulting, for example in a correlation between two consecutive days of 0.6583. 


\subsection{LIDAR dataset}
In addition to the motorcycle and Canadian weather datasets, we also analyzed the LIDAR (LIght Detection And Ranging) experiment dataset \citep{sigrist1994}, available in the \pck{HRW} \R package \citep{hrw2021}. This dataset was originally used in \cite{ruppert2003} to illustrate semiparametric regression and has more recently appeared in the functional data analysis literature \citep{meyer2024, goepp2025}. LIDAR is a remote-sensing technique widely applied to study atmospheric species. In particular, the DIAL (DIfferential Absorption Lidar) method uses two laser wavelengths: one tuned to an absorption line of the target species and the other slightly off-resonance. The ratio between the signals received from both laser sources can be used to estimate the concentration of the target species at a given range \citep{edner1992, svanberg2023}. 

The dataset analyzed in this section consists of 221 observations, each representing the logarithm of the ratio of received signals at frequencies on and off the resonance frequency of mercury. This dataset was collected during a study of atmospheric atomic mercury concentrations in an Italian geothermal field \citep{holst1996}.
We applied our method to the LIDAR dataset using five initial sets of cubic B-spline basis functions, with $K = 6, 10, 15, 20,$ and $30$. The optimal set was chosen using the generalized cross-validation (GCV) criterion, which yielded values of $0.0093$, $0.0081$, $0.0085$, $0.0096$, and $0.0176$, respectively. Based on these results, we selected the set with $K = 10$ as the optimal representation. The initial values for the parameters of our VEM algorithm and the hyperparameters were set as in the previous real-data applications.

For comparison, we also applied regression splines, smoothing splines, the Bayesian LASSO, and LASSO, all using the same set of $K = 10$ basis functions. Performance was competitive across methods in terms of adjusted $R^2$: smoothing splines ($0.9182$), regression splines ($0.9181$), LASSO ($0.9180$), Bayesian LASSO ($0.9162$), and our method ($0.9003$). Importantly, our approach further reduced complexity by selecting only the last 5 basis functions as relevant from the initial set. 

\begin{figure}[ht]
    \centering
    \includegraphics[width = 0.8\textwidth]{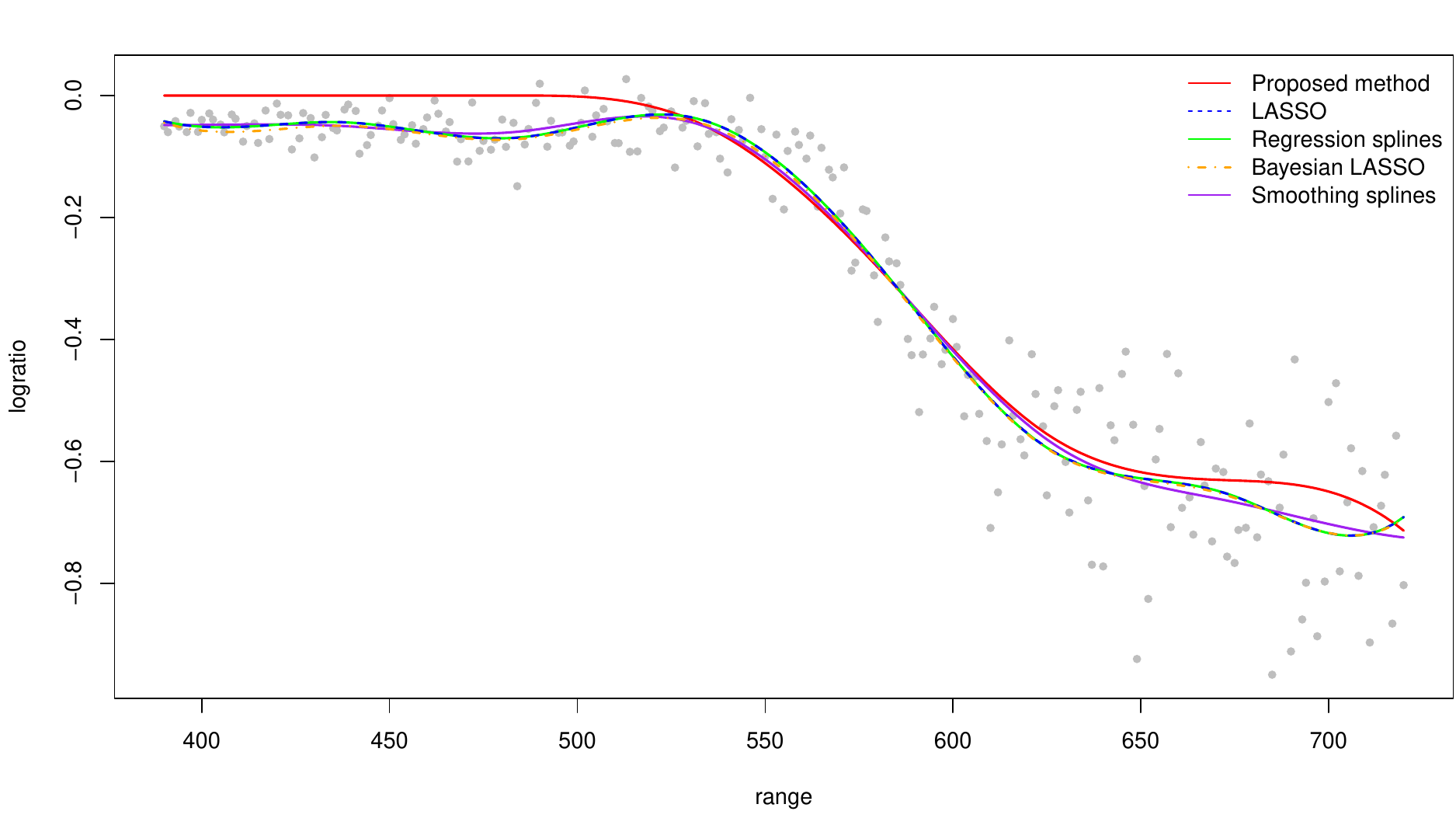}
    
    \caption{The $x$-axis represents the distance traveled before the light is reflected back to its source (and is referred to as the range), and the $y$-axis is the logarithm of the ratio of the signal received from two laser sources: one source has a frequency equal to the resonance frequency of mercury, and the other has a frequency off this resonance frequency. The estimated mean curves are plotted using our proposed method (red), regression splines (green), smoothing splines (purple), LASSO (blue), and Bayesian LASSO (orange).}
    \label{fig:data_lidar}
\end{figure}

Figure \ref{fig:data_lidar} shows the estimated curve for the log-ratio of the two laser signals over the range of distances. All methods captured the downturn between 520–580 meters. In particular, below 520 meters, our method was the only one to estimate the log-ratios as flat with value zero, whereas the estimates by the other methods oscillate around a value close to zero, but are not exactly zero or even flat. A log-ratio of zero at a given range indicates that both signals were returned with equal intensity, implying that no mercury was present at that range. The result found by our method is consistent with the concentration profile reported in \cite{holst1996}.

\clearpage
\section{Discussion}

In this work, we propose a variational EM algorithm to select basis functions for functional data representation in the presence of correlated errors. Simulation studies demonstrate that our method accurately identifies the correct set of basis functions from a candidate set while incorporating within-curve correlation across various scenarios. Additionally, our approach can be applied to different types of functional data, as evidenced by the satisfactory results obtained in the simulated studies using both B-splines and Fourier basis functions. Although our method does not quantify the uncertainty in estimating the correlation decay parameter, it effectively improves the measurement of uncertainty in functional data representation. 

When fitting repeated functional measurements (replicates), our method provides everything necessary to compute two functionals, one representing the group effect and another representing the individual effect. It is essential to highlight that these group and individual functional effects are not generated directly by our method but can be calculated by taking the functional mean of each group (group functional effect) and the difference between this group mean and the predicted curve for each individual (individual functional effect).

The application of our proposed method to the motorcycle and Canadian weather temperature datasets further showcased the practical utility of our method. Compared to the other techniques, our method achieved comparable or superior performance in terms of adjusted $R^2$, regardless of whether a within-curve correlation was present. Moreover, our method is flexible in handling data requiring fewer basis functions for a more parsimonious representation, as demonstrated with the motorcycle dataset, as well as scenarios needing more basis functions for accurate dataset representation, as seen with the temperature dataset. Our method effectively identifies and retains essential basis functions while selectively removing unnecessary ones. Additionally, our Bayesian approach allows the measurement of uncertainty via the construction of credible bands, which the other methods do not provide automatically.

Furthermore, we investigate the computational efficiency of a VB approach compared to a Gibbs sampler in performing basis selection through simulation studies. For this comparison, we assume independence among observations within a curve. The detailed results are provided in Section 3 in the Supplementary Material. Notably, our VB method demonstrated similar performance to the Gibbs sampler while using only thousandths of the computational time required by the Gibbs sampler, highlighting that variational inference methods provide an effective alternative for computing the posterior distribution, offering significant reductions in computational cost.

For future work, the within-curve correlation structure considered in this study can be extended to assume that both correlation and variance of the errors are functional. This would provide a more flexible dependence structure and improve the model's adaptability. Additionally, one could treat the number of basis functions,  $K$, as an unknown parameter with an associated prior distribution. This would require developing a VB algorithm, or a variation thereof, capable of accommodating changes in the latent space dimensionality across iterations.

\section*{Supplementary information}

Supplementary material is available in the PDF file ``Supplementary material.pdf".

\section*{Acknowledgments}
This research was financed by the Natural Sciences and Engineering Research Council of Canada, grant number RGPIN-2019-05915.

\section*{Declarations}

\begin{itemize}
\item The authors declare that they have no conflict of interest.
\item R code implementation is available at \url{https://github.com/acarolcruz/VB-Bases-Selection}
\end{itemize}

\bibliographystyle{apalike}
\bibliography{references_ssc}

\end{document}


\maketitle

\section{Other simulation results}

\subsection{Simulated Scenarios 1 and 2}
\label{supp:sim12}

\begin{figure}[ht]
    \centering
    \includegraphics[width = .30\textwidth, page = 1]{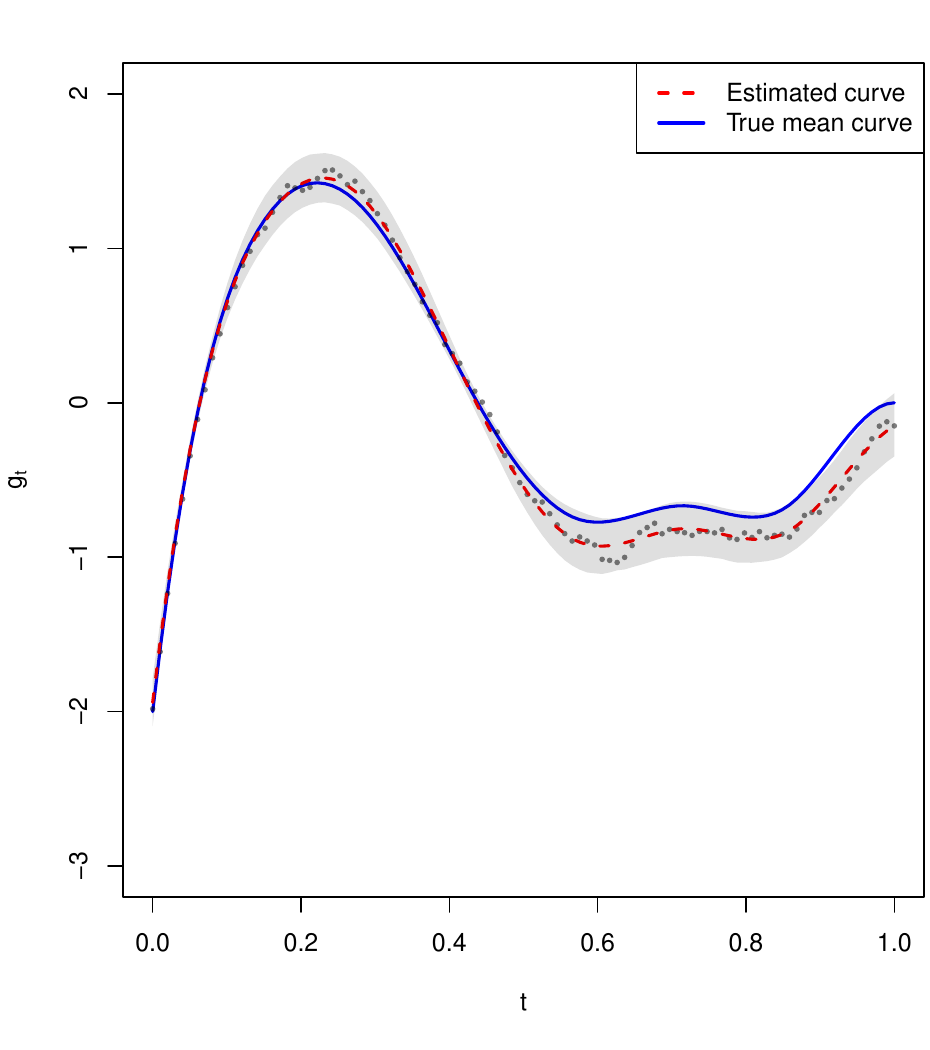}
    \includegraphics[width = .30\textwidth, page = 2]{Scenario1_alternative_plots_1_1.pdf}
    \includegraphics[width = .30\textwidth, page = 3]{Scenario1_alternative_plots_1_1.pdf}
    \includegraphics[width = .30\textwidth, page = 4]{Scenario1_alternative_plots_1_1.pdf}
    \includegraphics[width = .30\textwidth, page = 5]{Scenario1_alternative_plots_1_1.pdf}
    \caption{Estimated smooth curve for each of the five replicates for one simulated dataset (red) along with the true mean curve (blue), when $\sigma = 0.1$ and $w = 6$.}
    \label{fig:sim12_alternative_small_1}
\end{figure}

\begin{figure}[ht]
    \centering
    \includegraphics[width = .30\textwidth, page = 1]{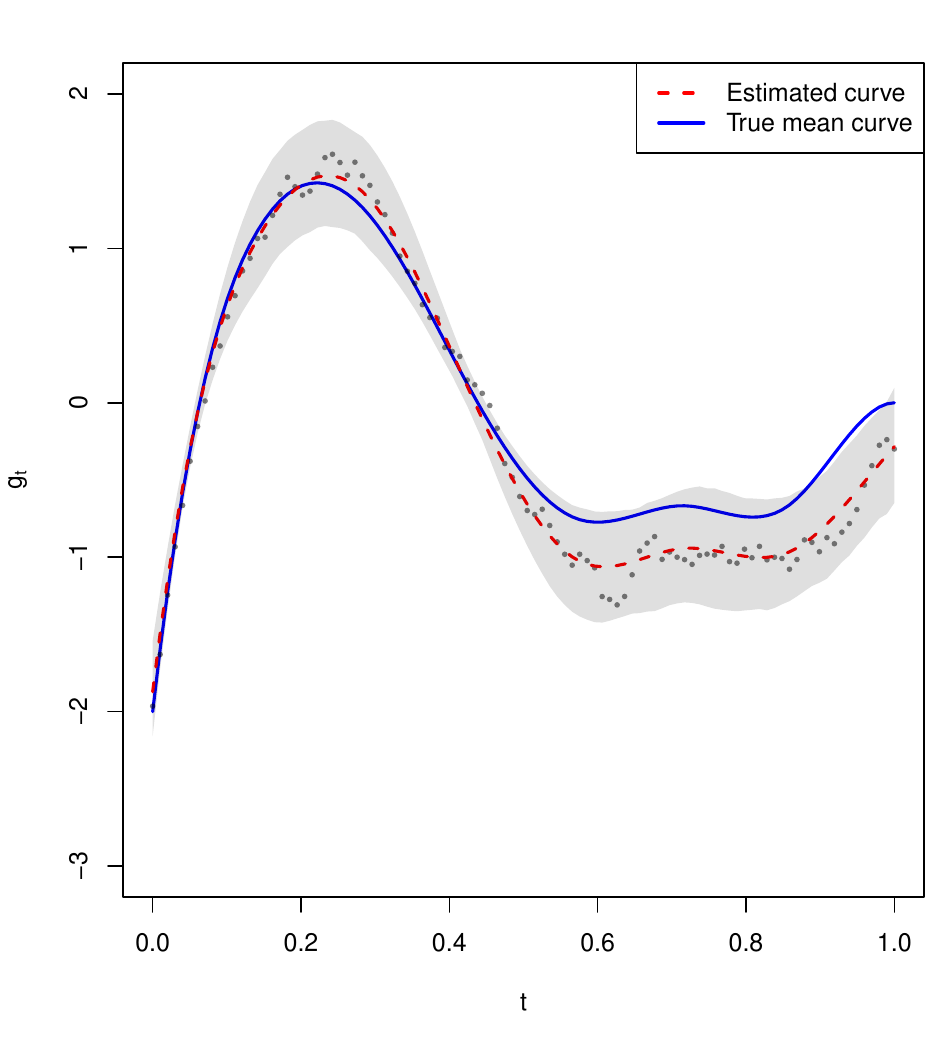}
    \includegraphics[width = .30\textwidth, page = 2]{Scenario1_alternative_plots_2_1.pdf}
    \includegraphics[width = .30\textwidth, page = 3]{Scenario1_alternative_plots_2_1.pdf}
    \includegraphics[width = .30\textwidth, page = 4]{Scenario1_alternative_plots_2_1.pdf}
    \includegraphics[width = .30\textwidth, page = 5]{Scenario1_alternative_plots_2_1.pdf}
    \caption{Estimated smooth curve for each of the five replicates for one simulated dataset (red) along with the true mean curve (blue), when $\sigma = 0.2$ and $w = 6$}
    \label{fig:sim12_alternative_large_1}
\end{figure}

In this section, we present an alternative way to show the results for the estimated curves from the simulated Scenarios 1 and 2 in the main text. For each scenario, we generate 100 datasets, each composed of $m = 5$ with $n_i = 100 \;\forall i$ with observed values equally spaced along the curves. To simplify the data generation, we generated all five curves in a simulated dataset using the same mean function, i.e., $g_i(t_{ij}) = g(t_{ij})$ for $i = 1, \dots, 5$. In practice, however, each curve may have a distinct mean function. Therefore, we would display the estimated smooth curve for each individual replicate as shown in Figures \ref{fig:sim12_alternative_small_1} and \ref{fig:sim12_alternative_large_1}.

\begin{figure}[ht]
    \centering
    \includegraphics[width = .30\textwidth, page = 1]{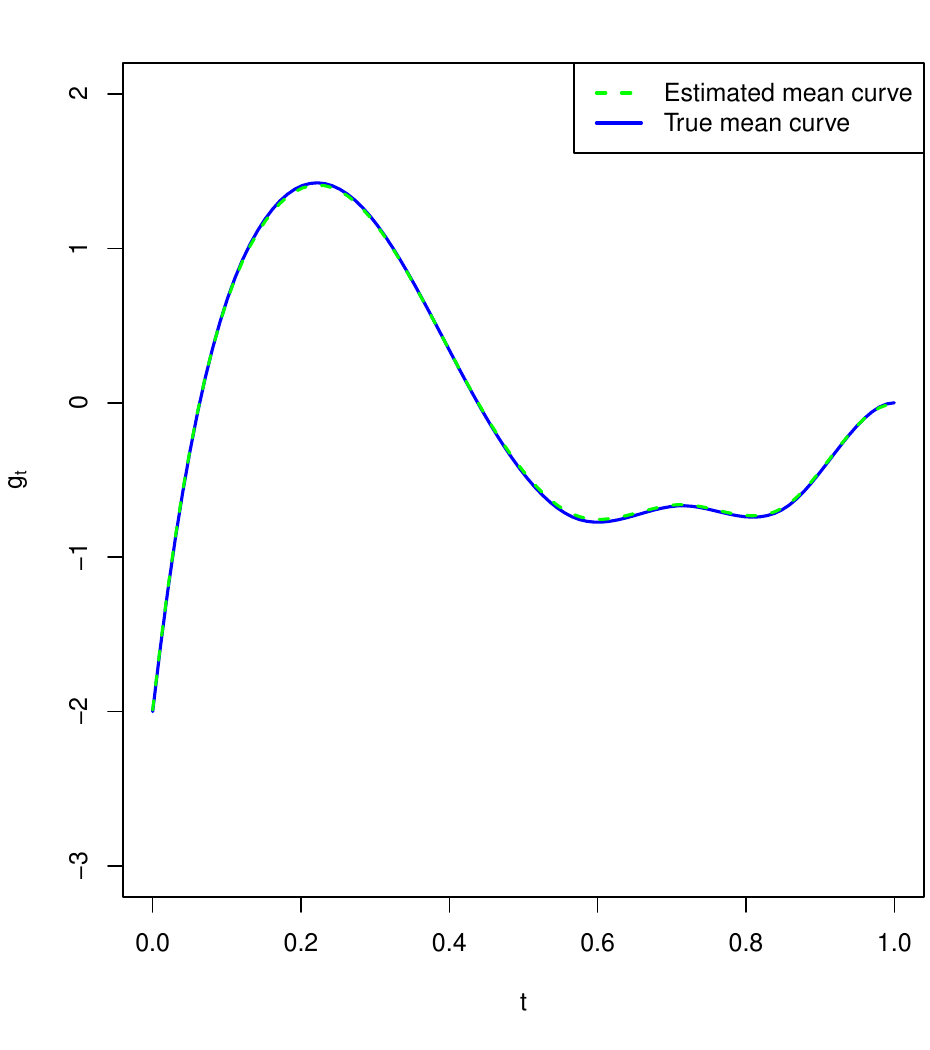}
    \includegraphics[width = .30\textwidth, page = 1]{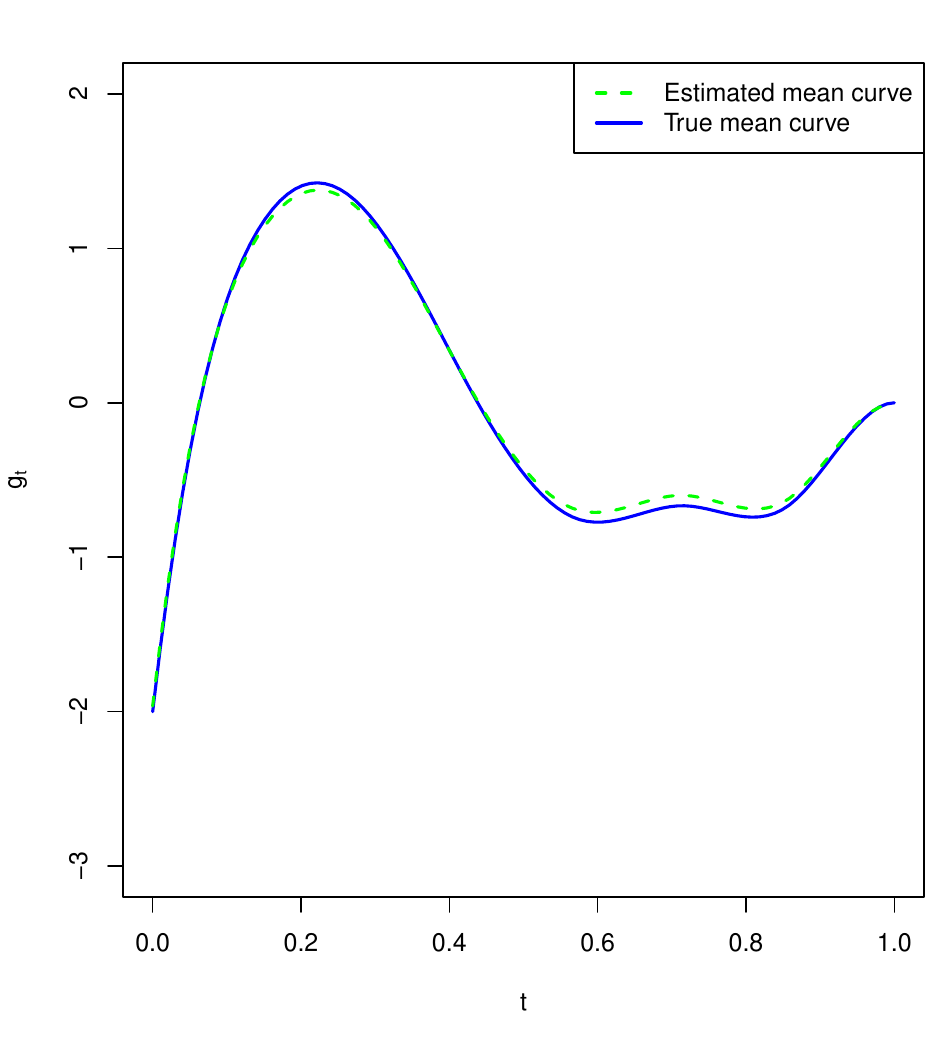}
     \caption{Estimated mean curve for one individual replicate across 100 simulated datasets (green) along with the true mean curve (blue), when $\sigma = 0.1$ (left panel), $\sigma = 0.2$ (right panel), and $w = 6$.}
     \label{fig:sim12_alternative_both}
\end{figure}

\clearpage
\section{Additional results for \pck{CanadianWeather} dataset}

\begin{figure}[ht]
    \centering
    \includegraphics[width = \textwidth, page = 1]{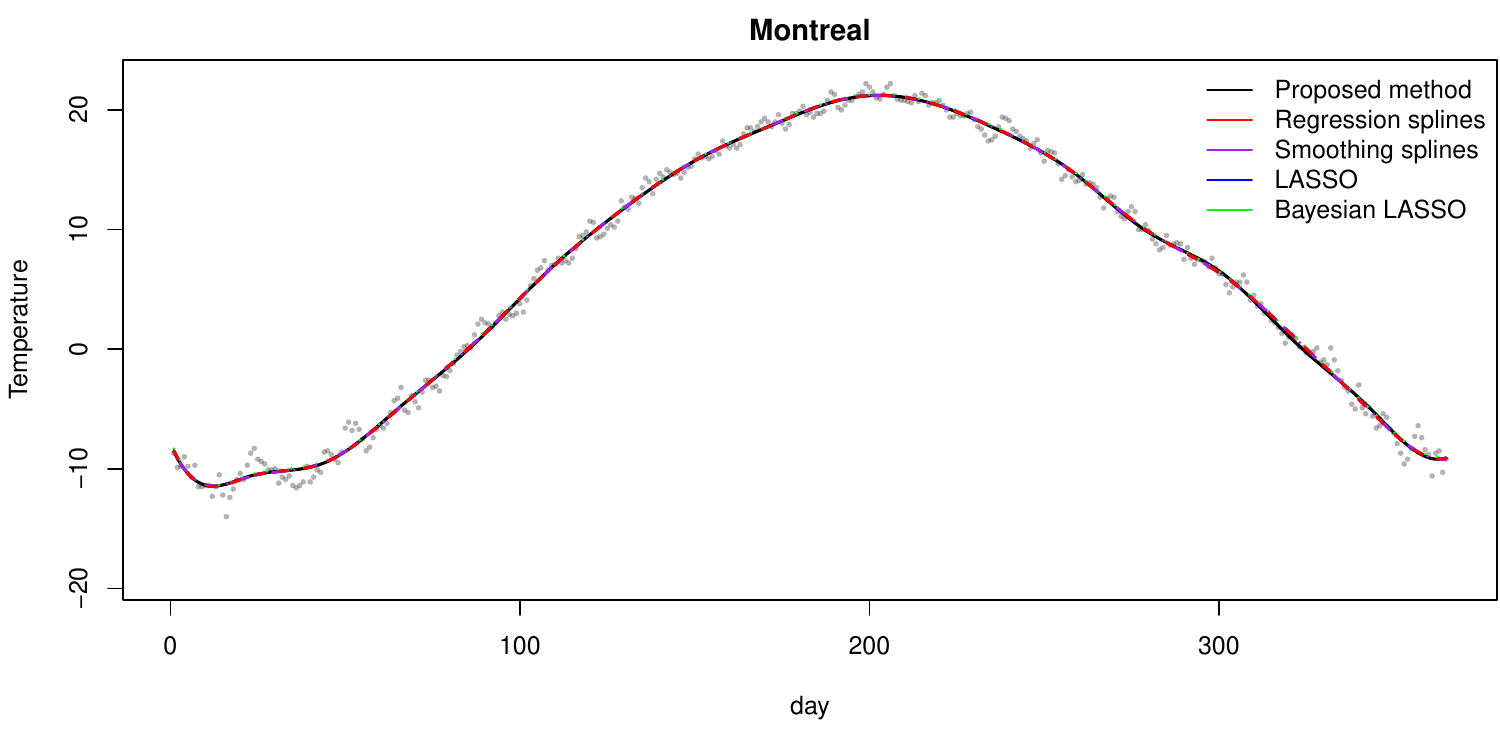}
    \includegraphics[width = \textwidth, page = 3]{plot_temp_all_1.pdf}
    \caption{Temperature analysis. Temperature in degrees Celsius in the $y$-axis and day of the year in the $x$-axis. Estimated mean curves for the daily temperature throughout the entire year for Montreal and Arvida using our proposed method (black), regression splines (red), regression splines (purple), LASSO (blue), and Bayesian LASSO (green).}
    \label{fig:data_temp_2cities1}
\end{figure}

\begin{figure}[ht]
    \centering
    \includegraphics[width = \textwidth, page = 2]{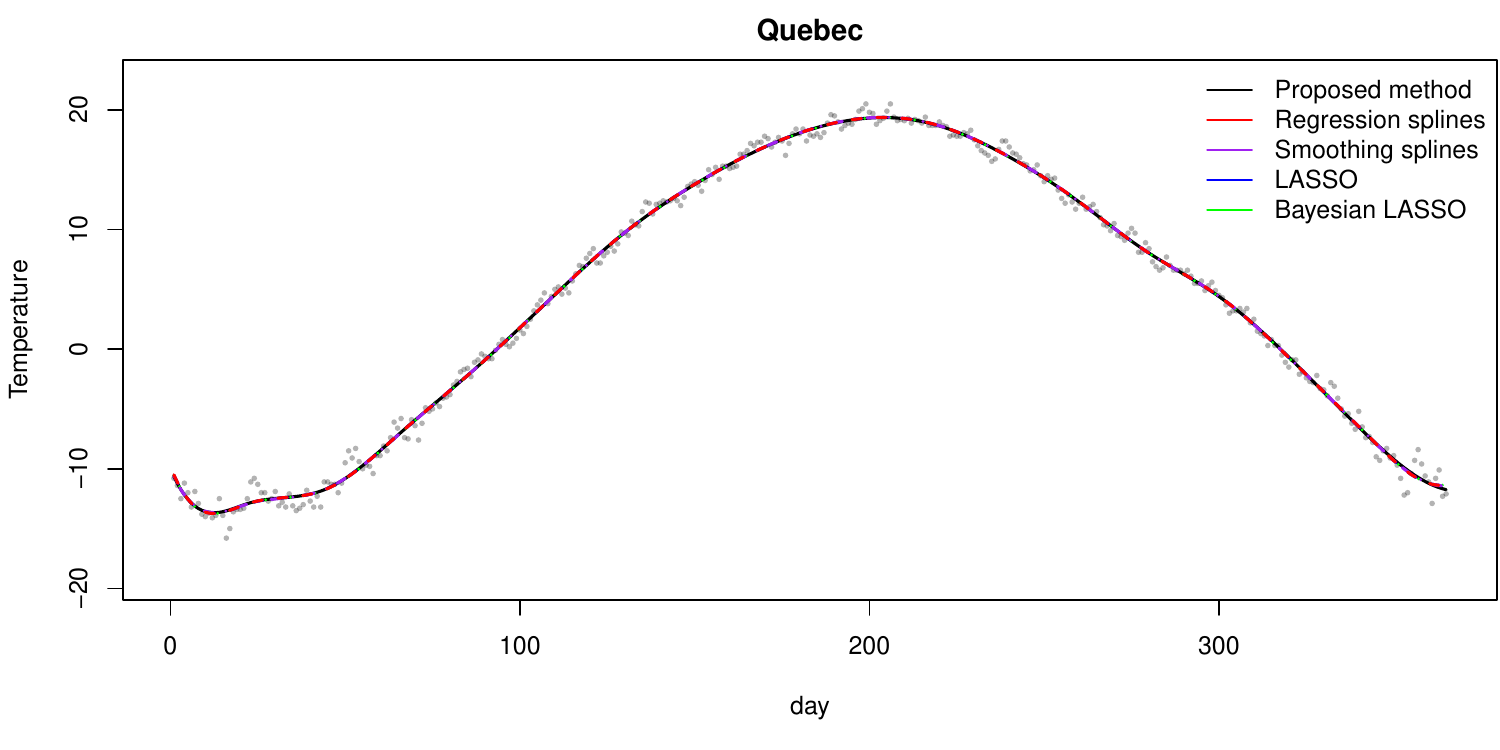}
    \includegraphics[width = \textwidth, page = 6]{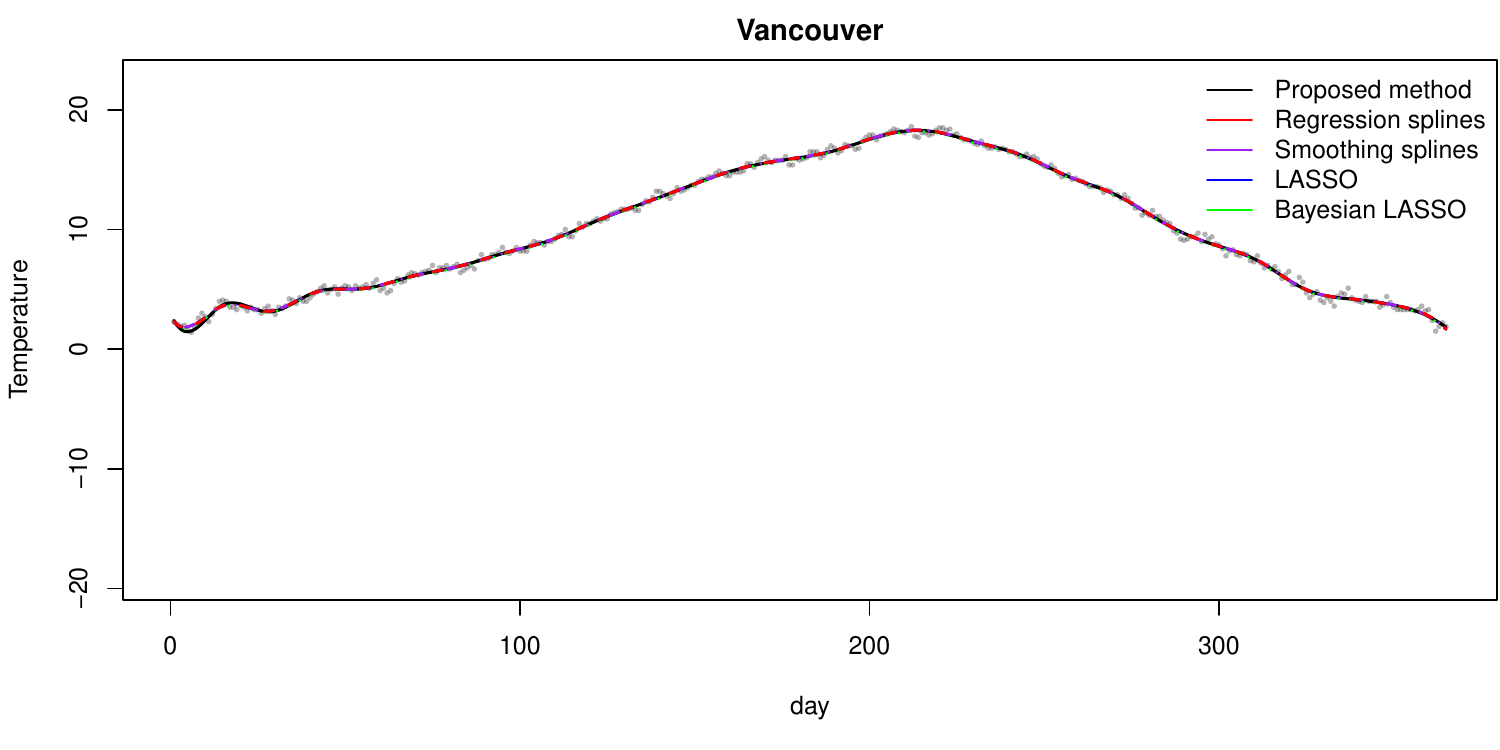}
    \caption{Temperature analysis. Temperature in degrees Celsius in the $y$-axis and day of the year in the $x$-axis. Estimated mean curves for the daily temperature throughout the entire year for Québec City and Vancouver using our proposed method (black), regression splines (red), regression splines (purple), LASSO (blue), and Bayesian LASSO (green).}
    \label{fig:data_temp_2cities1_2}
\end{figure}

In this section, we present additional results not shown in the main text for the \pck{CanadianWeather} dataset. Specifically, Figures \ref{fig:data_temp_2cities1} and \ref{fig:data_temp_2cities1_2} display the estimated mean curves for four cities (Montreal, Arvida, Quebec City and Vancouver) using each method, over the entire year. Table \ref{supptab:temp_gcv} shows the GCV values obtained from our proposed variational EM method for each of the six cities and each number of basis function $K$. Table \ref{supptab:res_temp_r2} presents the adjusted $R^2$ values obtained for each method considered in the analysis of the CanadianWeather dataset.

\begin{table}[ht]
\centering
\caption{Generalized cross validation (GCV) values obtained from our method for each of the six cities according to $K$.} 
\label{supptab:temp_gcv}
\begin{tabular}{l*{5}{r}}
  \toprule
City & \multicolumn{5}{c}{$K$} \\
\cline{2-6}
& 10 & 15 & 20 & 30 & 50  \\ 
\midrule
Montreal    & 0.0050 & 0.0048 & 0.0046 & 0.0049 & 0.0171 \\
Quebec      & 0.0041 & 0.0040 & 0.0038 & 0.0043 & 0.0181 \\
Arvida      & 0.0066 & 0.0062 & 0.0062 & 0.0060 & 0.0197 \\
Bagottville & 0.0047 & 0.0044 & 0.0043 & 0.0038 & 0.0135 \\
Sherbrooke  & 0.0076 & 0.0076 & 0.0071 & 0.0071 & 0.0165 \\
Vancouver   & 0.0061 & 0.0046 & 0.0046 & 0.0036 & 0.0295 \\
\bottomrule
\end{tabular}
\end{table}

\begin{table}[ht]
\centering
\caption{Adjusted $R^2$ values obtained from our method, regression splines, smoothing splines, Bayesian LASSO and LASSO. Optimal number of basis functions  for Arvida, Bagottville, Sherbrooke, and Vancouver was $K = 30$ and $K = 20$ for Quebec City and Montreal. The same set of basis functions was used across methods, based on the optimal number of basis function for each the city.} 
\label{supptab:res_temp_r2}
\begin{tabular}{lllllr}
  \toprule
City & Proposed method & Regression splines & Smoothing splines & Bayesian LASSO & LASSO  \\ 
  \midrule
Montreal    & 0.9956 & 0.9956 & 0.9956 & 0.9956 & 0.9956 \\
Quebec      & 0.9964 & 0.9964 & 0.9964 & 0.9964 & 0.9964 \\
Arvida      & 0.9944 & 0.9953 & 0.9952 & 0.9952 & 0.9952 \\
Bagottville & 0.9965 & 0.9967 & 0.9963 & 0.9966 & 0.9966 \\
Sherbrooke  & 0.9934 & 0.9940 & 0.9936 & 0.9938 & 0.9938 \\
Vancouver   & 0.9966 & 0.9969 & 0.9967 & 0.9968 & 0.9968 \\
   \bottomrule
\end{tabular}
\end{table}

\clearpage

\section{Comparative study between MCMC and VB}

In this section, we compare the computational efficiency of a VB algorithm to a Gibbs sampler in performing basis function selection, assuming independence among observations within a curve. As in the simulated Scenarios 1 and 2 in Section 3.1 in the main text, we generate five curves given by a linear combination of cubic B-splines with known true coefficients, plus some random noise generate by a normal distribution with mean of zero and variance of $\sigma^2 = (0.1^2, 0.5^2)$ for Scenarios 1 and 2 respectively. Thus, the curves are generated as follows: 

\begin{equation}
y(t_{ij}) = (-2, 0, 1.5, 1.5, 0, -1, -0.5, -1, 0, 0)'\mathbf{B}(t_{ij}) + \varepsilon(t_{ij}) ,
\end{equation}
where $\varepsilon(t_{ij}) \sim N(0, \sigma^2)$ and $\mathbf{B}(t_{ij})$ is the vector of the generated B-splines evaluated at $t_{ij}$. Furthermore, without loss of generality, we assume that the curves are observed in the interval $[0, 1]$ and at the same evaluation points, that is, $t_{ij} = t_j \; \forall i = 1, \dots, 5$

To compare both approaches we consider 100 simulated datasets where for the Gibbs sampler we utilize two chains each with 10000 iterations. We initialize both chains from different starting points:

\begin{itemize}
\item First chain: $\boldsymbol{\beta} = \mathbf{-1}, \boldsymbol{\theta} = \mathbf{\frac{1}{5}} , \sigma^2 = 1 $ and $\tau^2 = 1$ and
\item Second chain: $\boldsymbol{\beta} = \mathbf{1}, \boldsymbol{\theta} = \mathbf{\frac{4}{5}}, \sigma^2 = 5 $ and $\tau^2 = 5$.
\end{itemize}

Additionally, in the first chain, we initialize $\mathbf{Z}$ as one for all curves and basis functions, while in the second chain, we generate these values randomly from a Bernoulli distribution with probability of success 0.7. For the hyperparameters of the priors for $\sigma^2$ and $\tau^2$, we use non-informative priors, setting both hyperparameters in each prior to zero. To summarize the results, we first evaluate the convergence of the chains using the approach proposed by \cite{gelman1992}. Then, using a 50\% burn-in for each chain and thinning by 50, we obtain 100 sample points from each chain, resulting in 200 values to estimate the posterior of each parameter. The estimates of each parameter are computed as the maximum a posteriori estimate and the credible bands are constructed as in \cite{sousa2024}. For the VB algorithm, we initialize the required parameters and obtain their estimates as described in Section 3 of the main text.

\begin{figure}[ht]
 \centering
    \begin{subfigure}[b]{.49\textwidth}
        \centering
        \includegraphics[width = \textwidth]{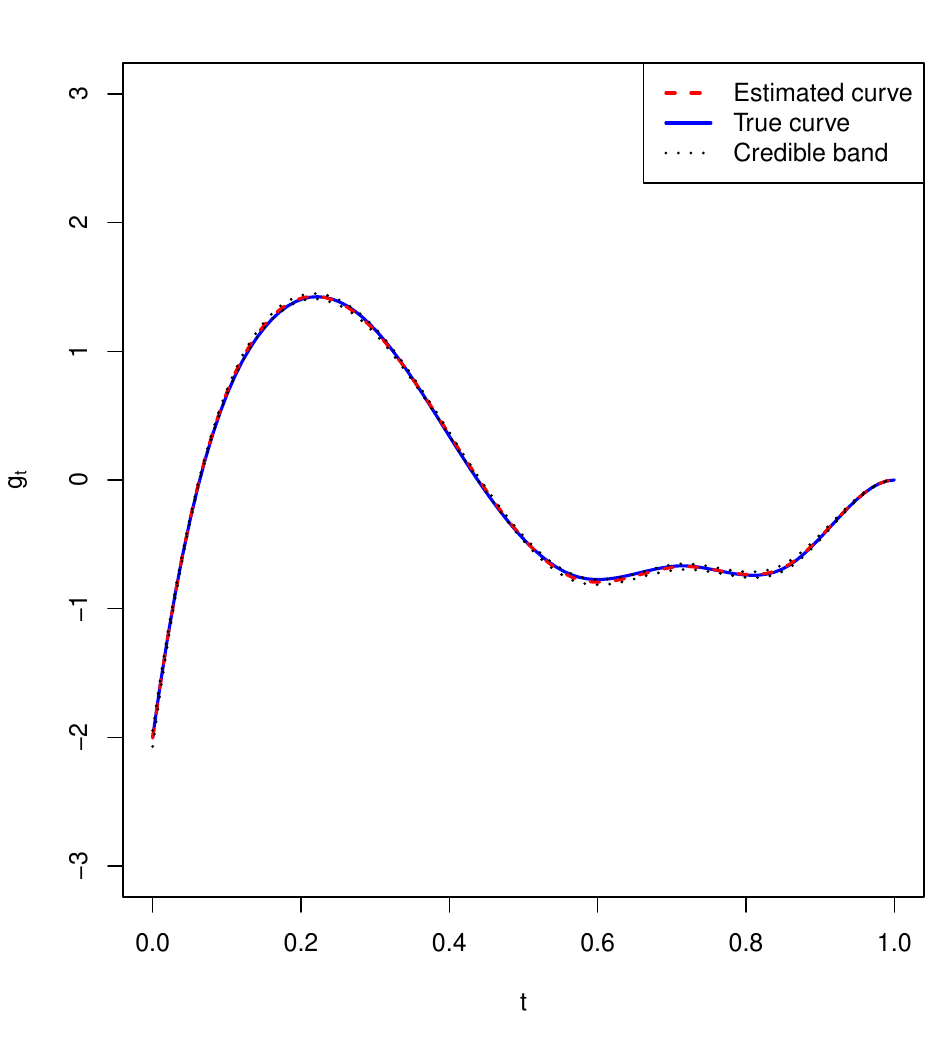}
        \caption{VB ($\sigma = 0.1$)}
    \end{subfigure}
     \begin{subfigure}[b]{.49\textwidth}
        \centering
        \includegraphics[width = \textwidth]{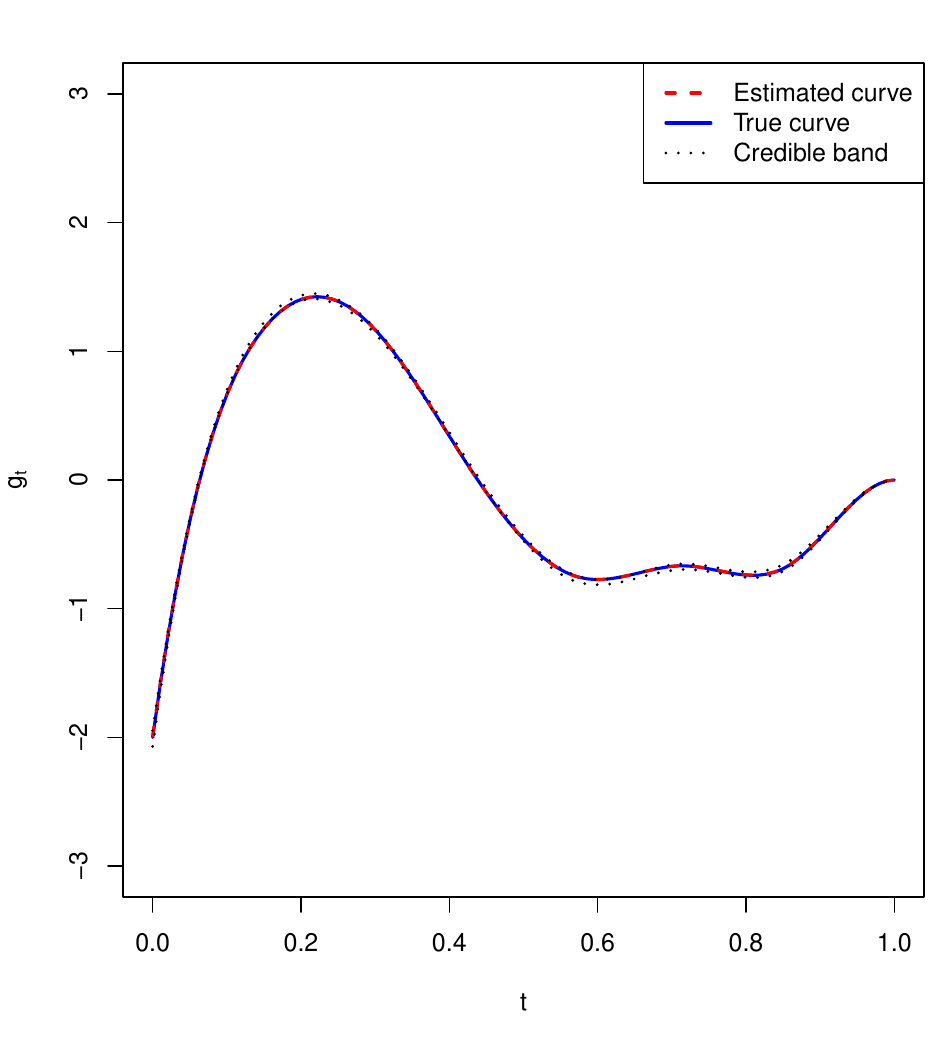}
        \caption{Gibbs sampler ($\sigma = 0.1$)}
    \end{subfigure}
    \begin{subfigure}[b]{.49\textwidth}
        \centering
        \includegraphics[width = \textwidth]{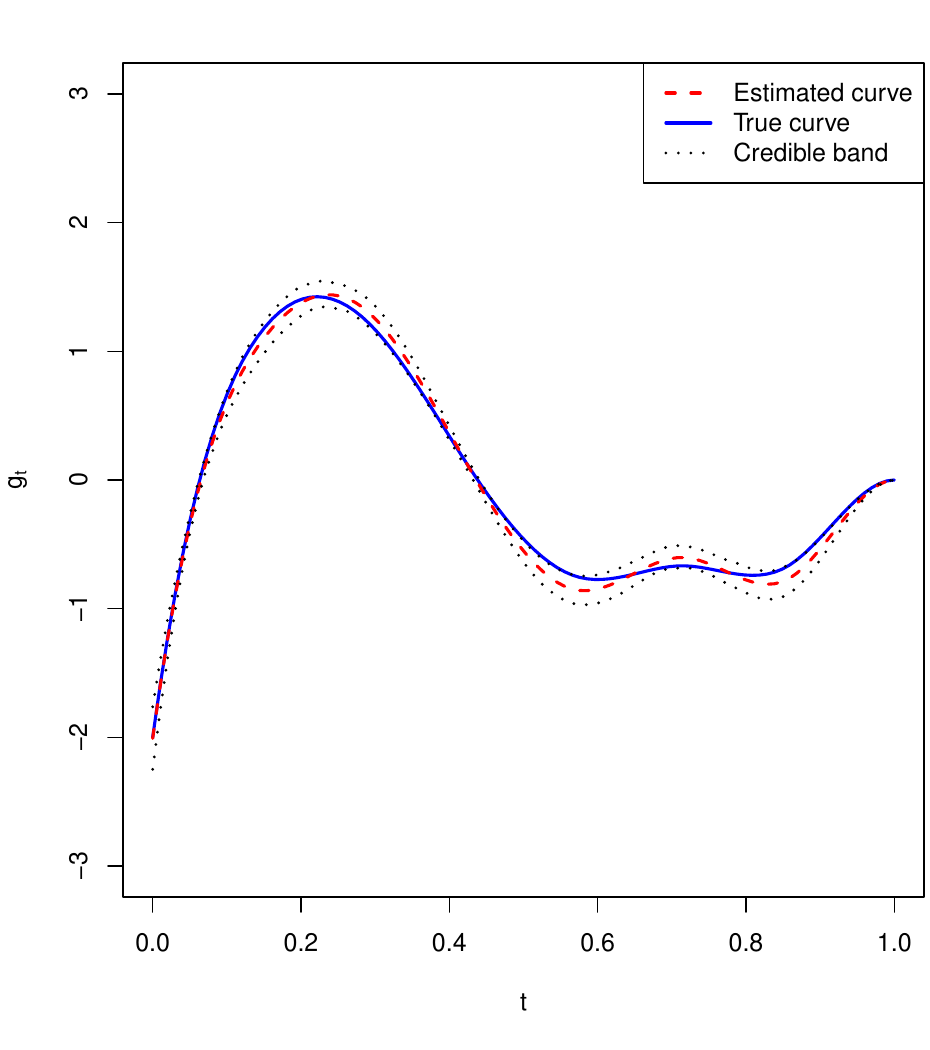}
        \caption{VB ($\sigma = 0.5$)}
    \end{subfigure}
    \begin{subfigure}[b]{.49\textwidth}
        \centering
        \includegraphics[width = \textwidth]{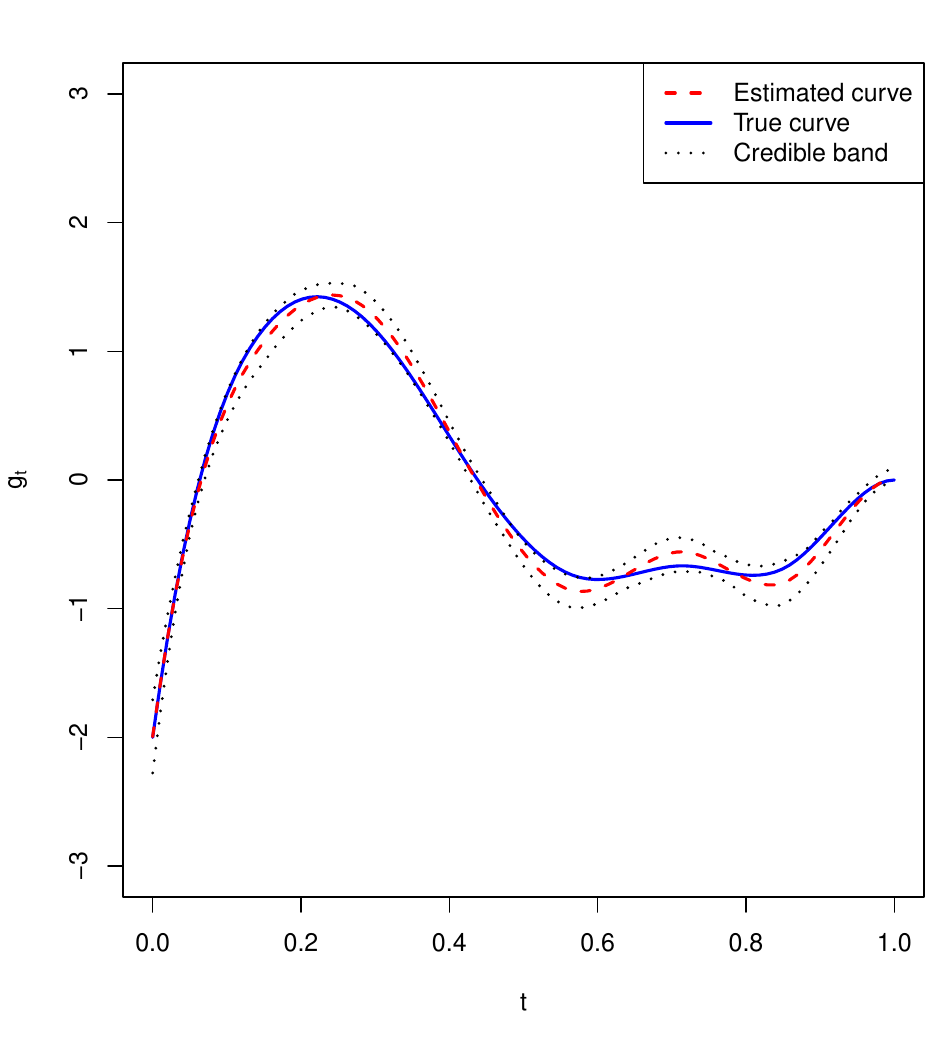}
        \caption{Gibbs sampler  ($\sigma = 0.5$)}
    \end{subfigure}
    \caption{True (blue) and estimated (red) mean curves with a 95\% equal-tailed credible band (dashed lines) for one simulated dataset accordingly to the data dispersion, $K = 10$ and $m = 5$. Figures (a) and (c) refer to the VB algorithm for $\sigma = 0.1$ and $\sigma = 0.5$ respectively. Figures (b) and (d) refer to the Gibbs sampler for $\sigma = 0.1$ and $\sigma = 0.5$ respectively.}
\label{fig:curve_sim_vb_gibbs}
\end{figure}

In Figure \ref{fig:curve_sim_vb_gibbs}, we present the estimated mean curve along with a $95\%$ equal-tailed credible band under both scenarios ($\sigma = 0.1$ and $\sigma = 0.5$) obtained from our VB method (left panels) and from the Gibbs sampler (right panels). Notably, for both approaches, the estimated mean curve closely aligned with the true curve consistently across the evaluation points and the credible bands for both methods are similar.  

\begin{figure}[ht]
    \centering
    \begin{subfigure}[b]{.49\textwidth}
        \centering
        \includegraphics[width = \textwidth]{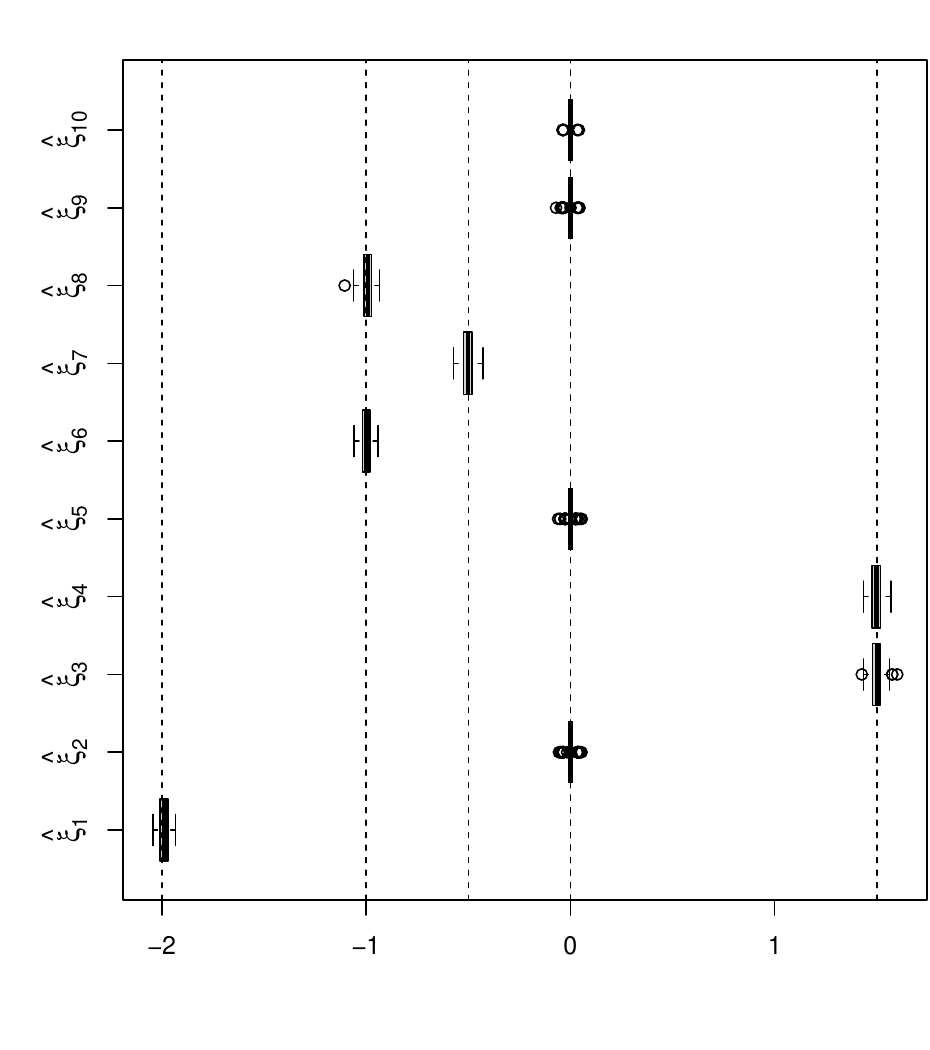}
        \caption{VB ($\sigma = 0.1$)}
    \end{subfigure}
     \begin{subfigure}[b]{.49\textwidth}
        \centering
        \includegraphics[width = \textwidth]{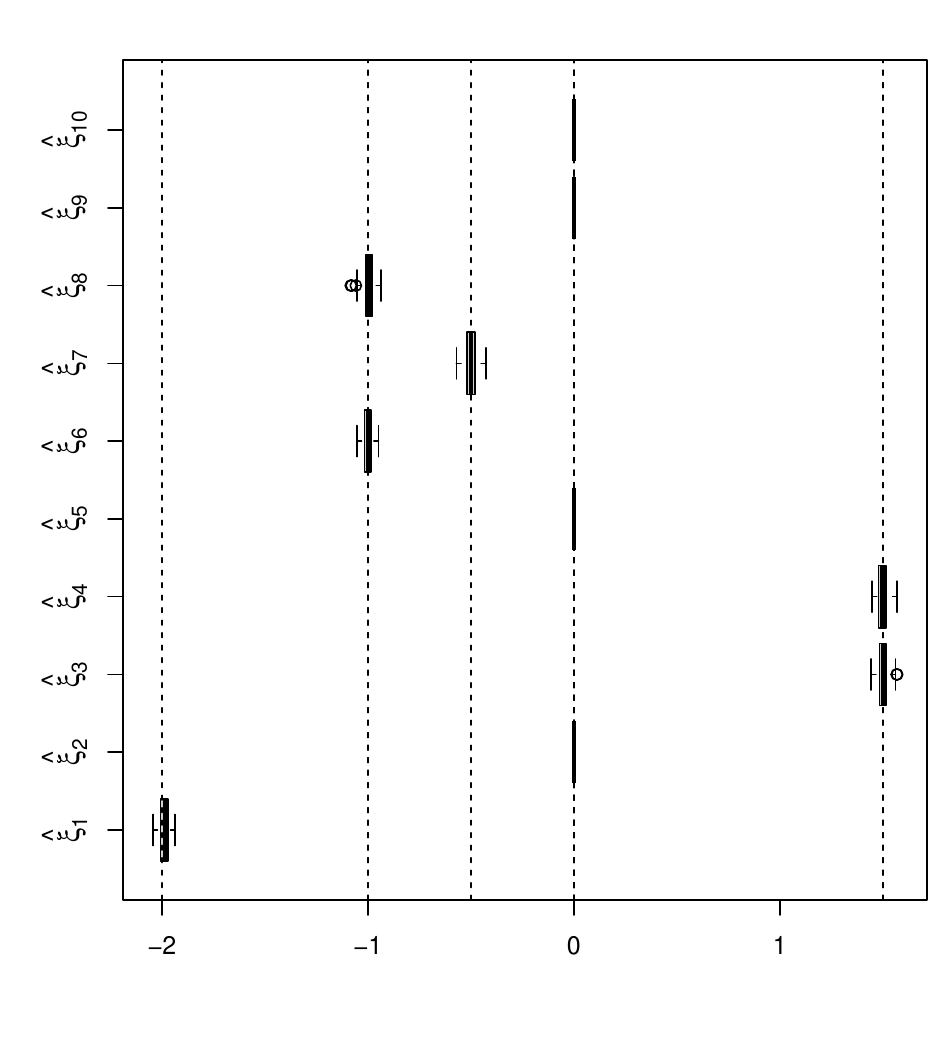}
        \caption{Gibbs sampler ($\sigma = 0.1$)}
    \end{subfigure}
    \begin{subfigure}[b]{.49\textwidth}
        \centering
        \includegraphics[width = \textwidth]{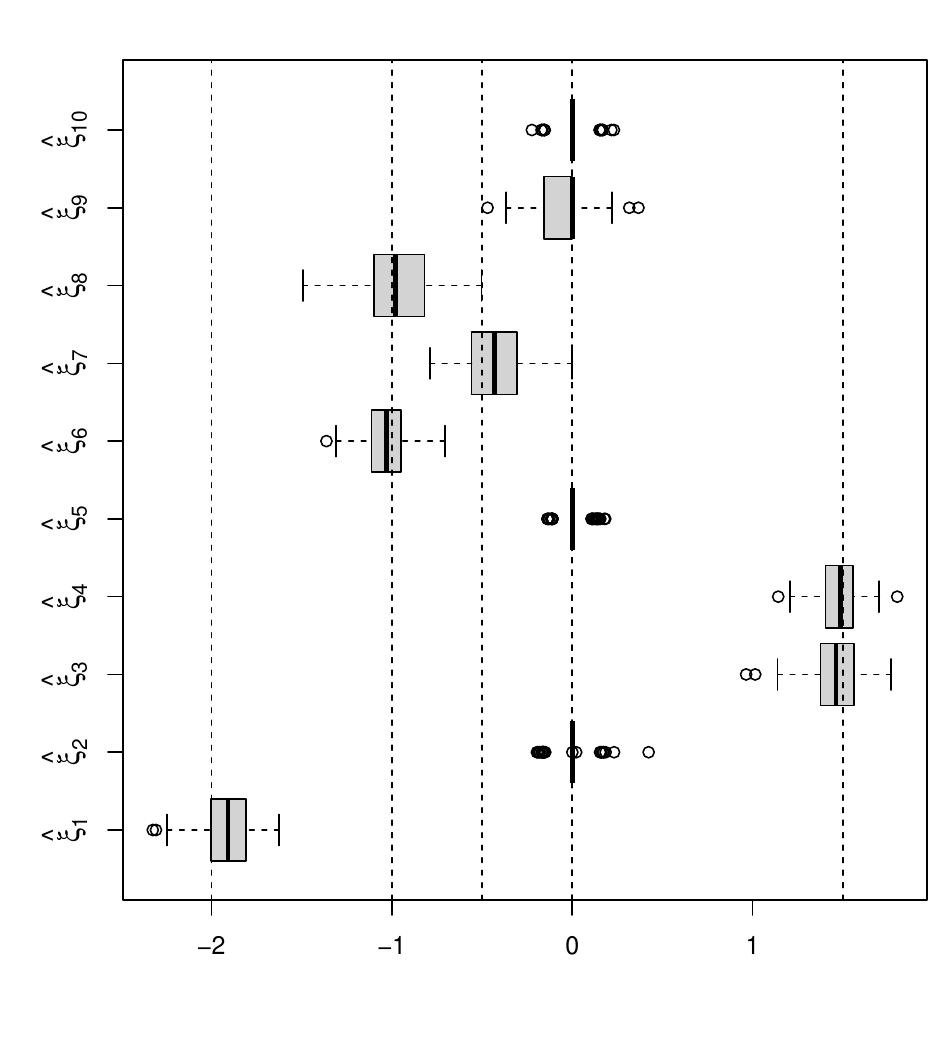}
        \caption{VB ($\sigma = 0.5$)}
    \end{subfigure}
    \begin{subfigure}[b]{.49\textwidth}
        \centering
        \includegraphics[width = \textwidth]{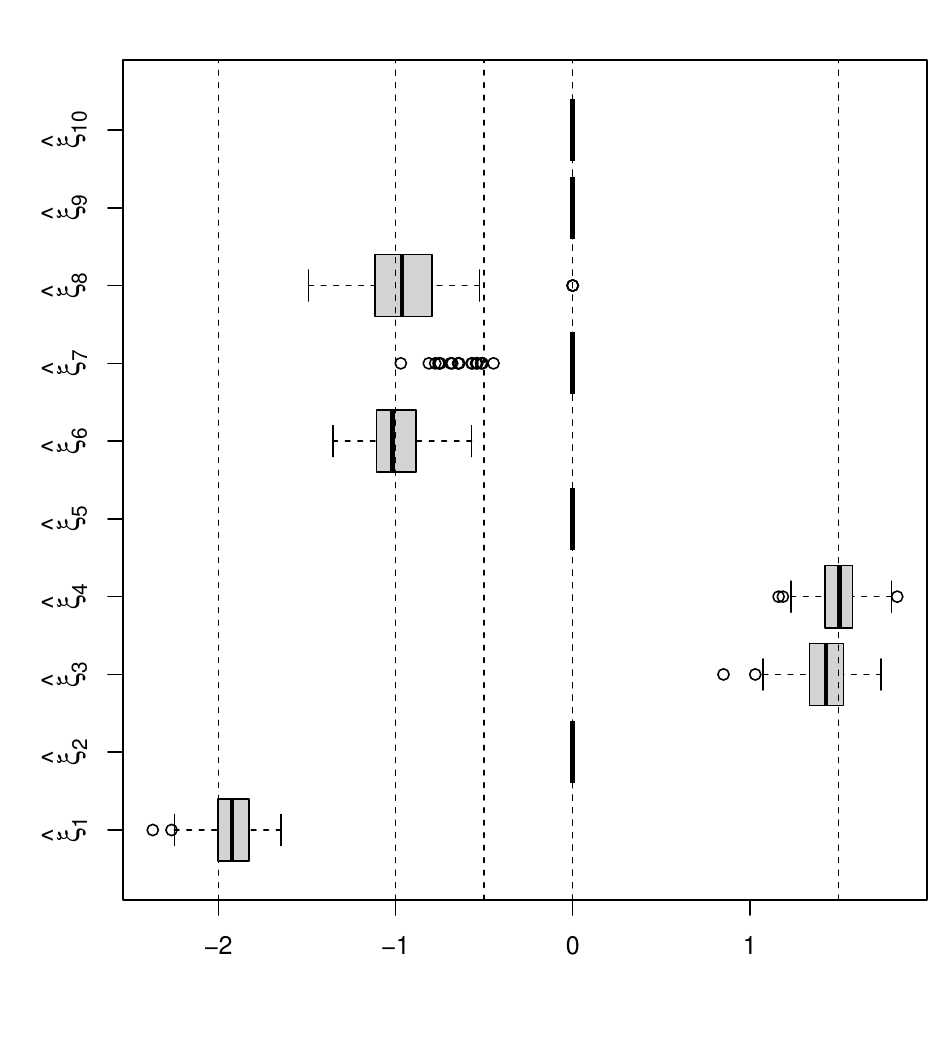}
        \caption{Gibbs sampler  ($\sigma = 0.5$)}
    \end{subfigure}
    \caption{Boxplots of the estimates of the basis function coefficients generated from 100 simulated datasets accordingly to the data dispersion, $K = 10$ and $m = 5$. The dashed lines correspond to the true values of the basis coefficients. Figures (a) and (c) refer to the VB approach for $\sigma = 0.1$ and $\sigma = 0.5$ respectively. Figures (b) and (b) refer to the Gibbs sampler for $\sigma = 0.1$ and $\sigma = 0.5$ respectively. }
\label{fig:boxplots_sim_vb_gibbs}
\end{figure}

Additionally, both models effectively selects the correct basis functions, as demonstrated in Figure \ref{fig:boxplots_sim_vb_gibbs}, where boxplots distant from zero correspond to true basis coefficients generated far from zero. 

We observe a strong consistency in the estimated posterior distributions between MCMC and VB. Both were successfully recover the true underlying structure in the data. Moreover, the VB method is in the order of thousands faster than MCMC, taking three seconds on average to obtain the posterior distributions. Therefore, this comparative study demonstrates the computational efficiency of VB in comparison with MCMC for performing basis function selection for multiple curves simultaneously.

\clearpage

\bibliographystyle{apalike}
\bibliography{references_ssc}